\documentclass[12pt,twoside]{article}

 \def\TheAuthor{S. Lounis}                                   
 \def\TheTitle{Theory of Scanning Tunneling Microscopy\footnotemark}           
\def\TheHeading{Theory of Scanning Tunneling Microscopy}                                      
 
\def\TheInstitute{Peter Gr\"unberg Institut \&
  Institute for Advanced Simulation}                  
\def\TheAddress{Forschungszentrum J\"ulich GmbH}              
 
\def\ThePart{C}
 
\def\TheChapter{6}
 
\usepackage{ferienschule_14}                                  
 
 \makeindex

\begin{document}

 \MakeTitel           
\tableofcontents     
 
\footnotetext{Lecture Notes of the $45^{{\rm th}}$ IFF Spring
 School
  ``Computing Solids - Models, ab initio methods and supercomputing''
  (Forschungszentrum J{\"{u}}lich, 2014). }
 
\newpage
 

\section{Introduction}

Scanning tunneling microscopy 
\iffindex{Scanning Tunneling Microscopy} (STM) is a tool that profoundly shaped
nanoscience and nano-technology. Since its  invention by Rohrer and
Binnig~\cite{Binnig81,Binnig82,Binnig83},  for which they were
awarded the 1986 Nobel prize in Physics,  STM experienced
revolutionary developments allowing to see for the first time 
nanostructures at the atomic scale. 
Another one is to access spintronics  at the
nanolevel. With increasing availability of low-temperature STM, local
electronic properties can be investigated with  unprecedented space
and energy resolution which opens the vista to completely new
applications. STM allowed the rather unique ability of accessing at the same time occupied 
and unoccupied electronic states.  In Fig.~\ref{chrono_stm} is shown a schematic view of the
chronological achievements of STM  during the last 20 years. Although
one cannot mention all important  milestones  in a single  figure,
Fig.~\ref{chrono_stm} tells us that after the initial application of STM
as a visualization tool of substrates  at the atomic level (surface
topography),  it developed quickly into a device for the manipulation
of atoms. Indeed in the 90's,  nanostructures such as corrals were
built atom by atom whereby a fundamental  quantum property, Friedel
oscillations induced by the presence of impurities in an  electron
gas, were observed and confined within man-made 
nano-objects~\cite{eigler,crommie,heller,manoharan}.  These
achievements were the prelude to functionalization of nanostructures
for different  applications, with the aim of  characterizing
and manipulating  not only the spin and charge of single atoms or
single molecules but also their position in much bigger
nanostructures.  

At the beginning of the $21^{st}$ century,
spin-polarized STM (SP-STM)  was 
invented~\cite{Wiesendanger90.1,Heinze2000} and applied
for the investigation of magnetic layers on different substrates. 
It was, for example, found that a manganese  monolayer 
deposited on tungsten(110) substrate is characterized by an 
antiferromagnetic ground state, confirming previous predictions made 
with first-principles calculations~\cite{bluegel},  and it was  nicely 
shown that contrary to the regular STM, the spin-polarized version
shows a  magnetic superstructure  on top of the chemical unit-cell.
Magnetic characterization is nowadays a routine work, that allowed the
discovery of new magnetic states, for example chirality, induced by
the existence of Dzyaloshinskii-Moriya interactions,  was revealed in
the magnetic contrast measured on a manganese layer deposited on
W(110) surface~\cite{heide} (See Chapter {\bf C 4} by S.~Bl\"ugel). Moreover, 
magnetism of finite nanostructures, nanowires for instance, on magnetic surfaces 
was recently characterized.~\cite{Loth2012,Lounis2013}

Being the ubiquitous apparatus for nanoscience, the range of phenomena
studied by STM is  continuously growing. Besides surface topography,
and the investigation of ground state  properties of excitations,
vibrational~\cite{ho1998}, magnetic~\cite{heinrich,heinrich2,
wulfhekel,khajetoorians} or optical~\cite{ho2003} properties, 
which allow chemical identification of  atoms, 
and even measurement of their magnetic anisotropy energy is today a
major topic studied  with state-of-the-art machines. Recently,
magnetometry measurements allowed to extract  quantitative values for
magnetic exchange interactions among adatoms separated by large
distances!~\cite{Wiebe1,Wiebe2}  Also other applications and developments of STM are
geared towards the  measurements of adhesion and strength of
individual chemical bonds, friction, studies of 
dielectric properties, contact charging, molecular manipulation and
many other phenomena from the micrometer down to subnanometer
scale. As Chen says in his  book~\cite{Chen93}: {\it It was often said
  that STM is to nanotechnology what the  telescope was to
  astronomy. Yet STM is capable of manipulating the objects it
  observes, to  build nanoscale structures never existed in Nature. No
  telescope  is capable of bringing Mars and Venus together.}

The actual playground of STM experiments was initially covered mainly
by theory. The advent  of such an instrument  urged the theoretical
community for  the development of new methods that allows the
understanding and prediction of phenomena  accessible with 
STM (see e.g. Refs.~\cite{Tersoff83.1,Tersoff85.1,Chen90.1,Chen90.2,
heinze2002} or Refs.~\cite{Lorente2000,Lorente2009,Persson,Fernandez,Fransson,
Hurley,
Lounis2010,Lounis2011} and 
many others). 
It is
clear that this tool will continue to experience further  evolutions
and to  play a pivotal role in further developments of
nanosciences. Thus more  challenges will be proposed to
theoreticians. It is not a surprise if after 20 years from its
inventions, several  books and reviews were dedicated to this
technique, to cite a few see Refs.~\cite{Guenterodt94,Wiesendanger94.1,Heinze_PhD,
Hofer2003,Bode2003,Wiesendanger2009,Houselt2010,Lorente2012}.

The goal of this lecture
is to review the basics behind the theory accompanying the experiment
which could be of interest for readers aiming to work in this field
or for those  who want to reassess some of the fundamental
concepts. Several flavors of the STM method have been  developed and
invented, we cannot go over all of them but we will discuss the
standard ones following partly the book of Chen~\cite{Chen93} and the lecture notes of Stefan Bl\"ugel~\cite{blugel_STM}.
\begin{figure}[ht!]
\begin{center}
\includegraphics*[width=.6\linewidth,trim=20mm 0mm 20mm
  0mm,angle=90]{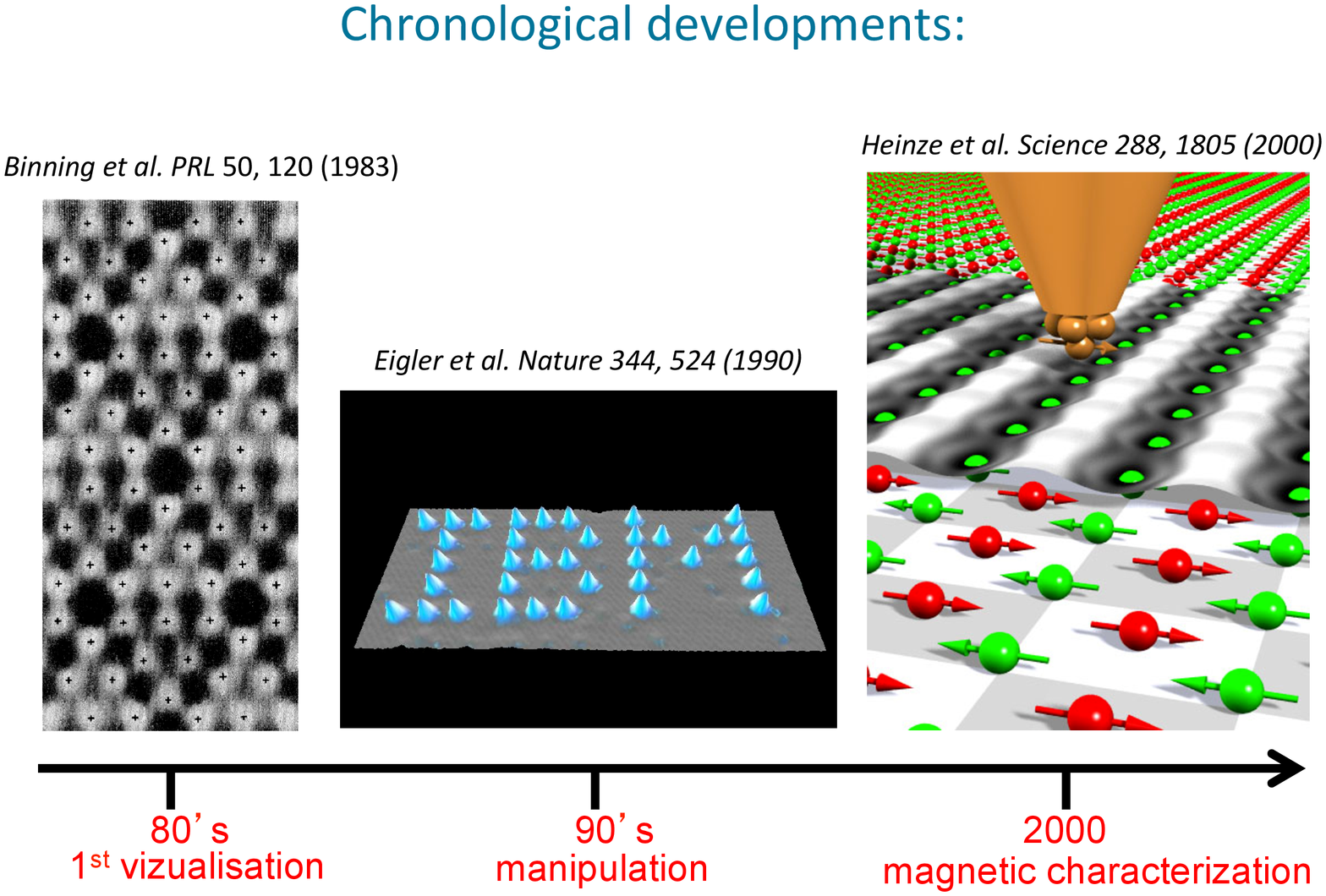}
\caption{Chronological developments of STM. At the early days of its
  invention, STM was used as a visualization tool for surface
  topography. It developed into a unique tool that allows access to occupied and unoccupied 
electronic states. Then  atomic manipulation was achieved and magnetic
  contrasts were realized thanks to the spin-polarized STM. We note
  that it is  impossible to mention all milestones that STM allowed
  to reach. }
\label{chrono_stm}
\end{center}
\end{figure}


\section{Description of scanning tunneling microscopy \label{description}}

Before discussing the basic theory explaining the measurements that
can be done with STM, a description of this tool is needed.  A crucial
ingredient in any STM is the probe tip that is attached to a
piezodrive, which consists of three mutually perpendicular
piezoelectric transducers (x, y and z piezo).  Upon applying a
voltage, a piezoelectric transducers contracts or expands which allows
to move the tip on the surface. Since the tip is not touching the
substrate, the flowing current, $I$, is weak and is obtained via a
tunneling mechanism through the vacuum. Fig.~\ref{stm_scheme}(a) shows
a schematic picture of an STM device.  If a bias voltage, $V$, is applied
between the tip and the sample, the tunneling current can change. 
\begin{figure}[ht!]
\begin{center}
\includegraphics*[width=.8\linewidth,trim=20mm 70mm 0mm
  30mm,angle=90]{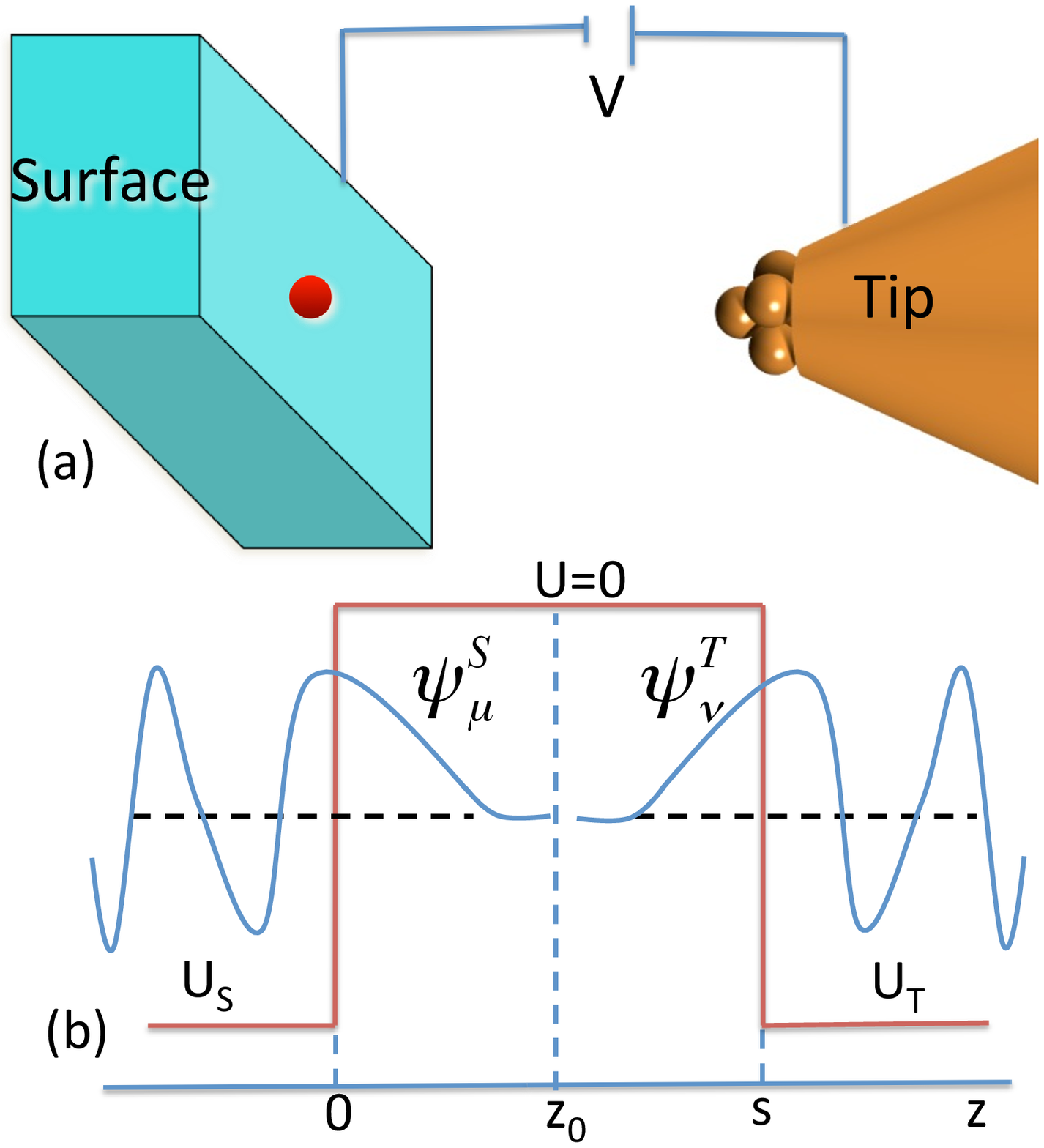}
\caption{A schematic picture of STM is shown in (a). A current can
  flow between the tip and the substrate through vacuum via a
  tunneling mechanism. In quantum mechanics, a particle has a non-zero
  probability of tunneling through a potential barrier which in the
  STM case is induced by the vacuum. A simple barrier as shown in  (b)
  explains the physics of tunneling. When the  two electrodes are far
  apart, the wave functions of both electrodes A and B decay  into the
  vacuum while the tunneling can take place if the electrodes are
  closer.  }
\label{stm_scheme}
\end{center}
\end{figure}

The simplest way to obtain a scanning tunneling microscope image is to
directly  measure the variation of the tunnel current as a function of
the scanning position while keeping the  distance between tip and
sample surface constant. A so-called current image is then obtained.
Instead of directly recording the atomic variation of the current,
however the usual procedure is to keep the tunnel  current constant
while scanning over the surface. This is done by changing the distance
between the tip and surface using  a feedback loop. In order to get an
image, the voltage  required at the piezoelectric crystal to adjust
the distance is recorded. One obtains then the so-called  constant-current
STM image. A further operation mode is the spectroscopy acquisition by
STM. It is usually done by interrupting the feedback in order  to keep
for the $I$--$V$ spectroscopy data acquisition the tip-sample
separation constant. This can be done at any  desired surface spot or
for every pixel in a STM image. An extended discussion of the
different operating modes will be given once the theory of STM is
presented.

Obviously, if one wants to understand the working mechanism of STM and
simulate the  experiment, one could think of simulating the whole
setup, i.e. considering  a tip, a substrate, a bias voltage and
calculate the tunneling current or conductance (see e.g. Chapter {\bf A 9} by S.~Tsukamoto).  
Although actual
ab-initio methods are capable of handling few hundreds up to few
thousands of atoms in a unit-cell (see e.g.  Chapter {\bf D 6} by R.~Zeller), technical
issues can occur. For instance, a periodic supercell approach would
lead to a  non-realistic scenario of multiples tips scanning the
substrate at the same time. Methods based on Green functions allow to
consider two perfect semi-infinite substrates separated by vacuum. One
of the substrates would simulate the surface,  on the other substrate,
a model tip can be embedded (see e.g. Ref.~\cite{Ernst}).   Although this scheme is appealing, one
would be facing the problem of choosing the right model  for the tip,
which is far from being easily accessible experimentally. All of those  arguments
stimulated approximations that are very often used successfully for the
understanding of STM-experiments 
but they also bear limitations.

\section{{The concept of tunneling}}
Here we describe briefly elementary theories of tunneling through a
one-dimensional  potential barrier, which will help us to grasp the
basic concept used in STM. 
In quantum mechanics, the electron feeling a potential U(z), for
example the one shown in Fig.~\ref{stm_scheme}(b) considering 
$U_S =U_T
= -U$,   is described by a wave function $\psi(z)$, which satisfies
the Schr\"odinger equation,
\begin{equation}
-\frac{\hbar^2}{2m}\frac{d^2}{dz^2}\psi(z) +U(z)\psi(z) = E\psi(z),
\label{eq.1}
\end{equation}
at a given position $z$. In this elementary model, the STM setup is simplified
to a one-dimensional potential barrier where  the vacuum is modeled
by the potential barrier $U$, while its left and right sides shown in
Fig.~\ref{stm_scheme}(b) represent  the substrate, $S$, and the tip, $T$.   
When $E > |U|$, the solutions of Eq.~\ref{eq.1} are $\psi(z)=\psi(0)e^{\pm ikz}$
where $k = \frac{\sqrt{2m(E-|U|)}}{\hbar}$ is the wave vector.  In the
classically forbidden region, within the barrier, the solution is
given by $\psi(z) = \psi(0) e^{-\varkappa z}$  with
$\varkappa=\frac{\sqrt{2m(|U|-E)}}{\hbar}$ being the decay constant that
describes an electron penetrating through the barrier into the $+z$
direction. The probability  density for the observation of an electron
near a point $z$ is  finite in the barrier region and is proportional
to $|\psi(0)|^2e^{-2\varkappa z}$. Additionally, electrons  can propagate in the
opposite direction ($-z$-direction) indicating  that tunneling is
bidirectional. 

The total wave function in every region, sample, barrier and tip are written as:
\begin{eqnarray}
\psi_S&=&e^{ikz} + A e^{-ikz} \\
\psi_{\mathrm{Barrier}}&=&Be^{-\varkappa z} + C e^{\varkappa z} \\
\psi_{T} &=& D e^{ikz}
\end{eqnarray}
The coefficients $A$, $B$, $C$ and $D$  take care of the reflection and transmission 
of the electrons, they are obtained by matching of the wave 
functions and their derivatives $d\psi/dz$ at the two interfaces, 
sample--barrier and barrier--tip. The incident 
current density $I_i= \hbar k/m$ and the transmitted current $I_t$
\begin{equation}
I_t = -i\frac{\hbar}{2m}\left(\psi_T^*(z)\frac{d\psi_T(z)}{dz} -
\psi_T(z)\frac{d\psi_T^*(z)}{dz} \right) =\frac{\hbar k}{m}|D|^2
\label{current-density}
\end{equation}
define the barrier transmission coefficient $T$ which is given by the ratio between the transmitted 
current density and the incident current density:

\begin{equation}
T=\frac{I_t}{I_i} =|D|^2 =\frac{1}{1+\frac{(k^2+\varkappa^2)^2}{4k^2\varkappa^2}\cdot \sinh^2(\varkappa s)}
\end{equation}
which simplifies in the limit of a strongly attenuating barrier (large decay constant $\varkappa$)
\begin{equation}
T\sim\frac{16\varkappa^2k^2}{(k^2+\varkappa^2)^2}\cdot e^{-2\varkappa s}
\label{eq.T}
\end{equation}
where $s$ defines the location of electrode $T$ (Fig.~\ref{stm_scheme}(b)).

From this basic model, some important features of a more realistic
metal-vacuum-metal  tunneling can be explained. Let us first evaluate the decay constant magnitude  which is defined 
defined by the work function $\Phi$ primarily if the electrons involved 
in the tunneling process are lying close to the Fermi energy of both electrodes. 
Indeed $\Phi$ is defined by the minimum energy
required to remove an electron from the bulk to the vacuum level. In
general, the work function  depends not only on the material, but also
on the crystallographic orientation of  the surface but to simplify
our discussion we assume it to be the same for the tip and sample
($\Phi_S = \Phi_T = \Phi$). In our model  $|U_S|$ and
$|U_T|$ are respectively replaced by their respective work functions. 
Here the decay constant $\varkappa =\frac{\sqrt{2m\Phi}}{\hbar}$ is of the order 
of $\sim 1$\AA$^{-1}$ for the typical values of the work function ($\sim 5 eV$).
 The typical values of
work functions of materials used in STM experiments, together with the
decay constants,  are listed in Table~\ref{work-function}. According to
Eq.~\ref{eq.T}, the current decays by one order of magnitude per 1\AA.

Even though this model is too simple to describe realistic STM
experiments it explains the high sensitivity to height changes in the
sample topography. Also it demonstrates that during tunneling, the tip's atom, 
that is the closest 
to the substrate, is the main atom involved in the tunneling process!


\begin{table}
\begin{center}
\begin{tabular}{lrrrrrrrr}
\hline Element & Al & Au & Cu & Ir & Ni & Pt & Si & W \\ \hline
$\Phi(eV)$ & 4.1 & 5.4 & 4.6 & 5.6 & 5.2 & 5.7 & 4.8 & 4.8 \\ $\varkappa
($\AA$^{-1})$& 1.03 & 1.19 & 1.09 & 1.21 & 1.16 & 1.22 & 1.12 & 1.12 \\ \hline
\end{tabular}
\caption{Work functions and decay constants according to Ref.~\cite{Chen93} for selected materials.}
\label{work-function}
\end{center}
\end{table}

From this simple one dimensional model one can derive the principle
used by Binning and Rohrer when they invented the STM. Their argument
to explain the  ability of  STM to achieve large lateral resolutions
and by that probe the electronic structures of various materials at an
atomic scale ($\sim 2$\AA) is: because of the tunneling through
vacuum, a large lateral resolution much smaller than the radius of the
tip-end, $R$, is possible  if the distance between the tip-end and the sample
surface, $\Delta z$, is much smaller than the tip radius~\cite{Quate1986}. Near the tip-end, the
current lines are almost perpendicular  to the sample surface (Fig.~\ref{lateral_stm}). 
At a
point $\Delta x$ on the tip, the distance to the sample surface, $z$, 
is increased by
\begin{equation}
\Delta z \sim \frac{\Delta x^2}{2R}.
\end{equation}
Assuming that at each point the tunneling current density follows the
formula for the one-dimensional case, Eq.~\ref{eq.T}, the lateral
current distribution is 
\begin{equation}
I(\Delta x)\sim e^{-2\varkappa \frac{\Delta x^2}{2R}}.
\end{equation}
Typically, $\varkappa \sim 1$\AA$^{-1}$. For $R = 10$\AA, at $\Delta x \sim
4.5$\AA, the current drops by a factor of $\sim e^{-2}$,  that is about one
order of magnitude.  The diameter of such a current column is the
resolution limit, which is about 9\AA. Therefore with moderate
means,  a very high lateral resolution is expected.  Nowadays,
achievements of STM largely exceeds this expectation. 

\begin{figure}[ht!]
\begin{center}
\includegraphics*[width=.6\linewidth,angle=0]{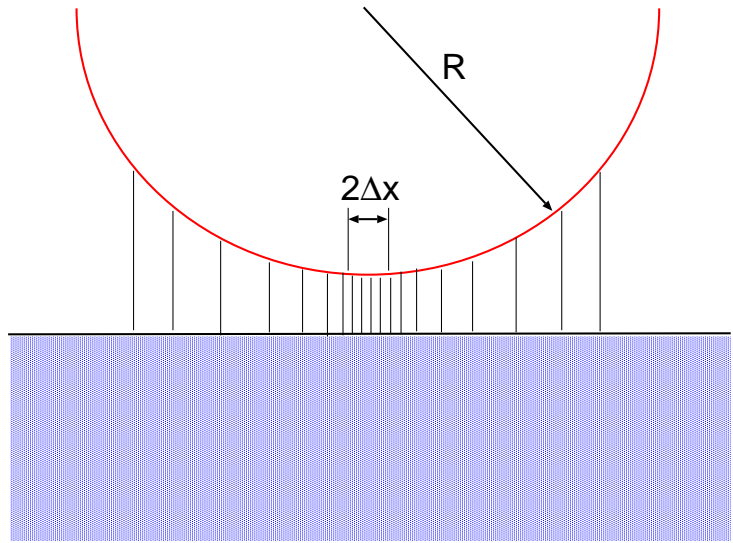}
\caption{Estimation of the lateral resolution in STM. Out of a
  spherical tip model with  radius $R$ very close to the surface, the
  lateral resolution of STM can  be evaluated. The tunneling current
  is concentrated at the vicinity of the closest point to the
  substrate.}
\label{lateral_stm}
\end{center}
\end{figure}



\section{Modeling currents}
\subsection{Bardeen's approach}\iffindex{Bardeen's Approach}
\subsubsection*{ One-dimensional case \label{one-dim}}

The planar tunneling junction problem treated by Bardeen is
schematically shown in  Fig.~\ref{stm_scheme}(b). The model used by
Bardeen, called also the transfer Hamiltonian method~\cite{Bardeen},  
and extended
later on by Tersoff and Hamann~\cite{Tersoff83.1,Tersoff85.1} 
and Chen~\cite{Chen90.1,Chen90.2} to STM, has naturally
limitations by its assumptions but gives a fundamental understanding
on the ability of STM  to reach high space and energy resolution. Here
are some assumptions that are assumed in the original derivation of
the  Bardeen's approach:

First of all, the electron tunneling is treated as a one-particle
process, i.e.  the mutual interaction between electrons during
tunneling  is neglected which is a reasonable approximation in the low
tunneling regime.   Furthermore, a direct interaction of tip and
sample resulting in the  formation of coupled electronic states is not
taken into account. This assumption  is valid if the tip-sample
distance is large enough, i.e.  a distance larger than $\sim$
4\AA~should be sufficient. Note that in our discussion  elastic tunneling
is assumed, i.e. no energy loss of the  electrons with quasi-particles
of the electrodes, e.g. plasmons, phonons,  spin-excitations is
considered in this section. Of course, recently models taking  care of this kind of
interactions were developed (see Chapter {\bf C 7} by M. Wegewjis 
for detailed examples) and 
a simple discussion on the treatment of inelastic tunneling is given later on in this 
manuscript. 

When the two electrodes, representing the tip and the sample,  are far
apart,  the wave functions  of electrode $S$, representing the
unperturbed substrate, or of the  unperturbed electrode $T$,
representing the tip, satisfy  the Schr\"odinger equation of the free
electrode  $S$ or $T$,
\begin{equation}
i\hbar \frac{\partial \psi^i}{\partial t} =
\left(-\frac{\hbar^2}{2m}\frac{\partial^2}{\partial z^2} +
U_i\right)\psi^i,
\label{eq.bardeen.1}
\end{equation}
where $U_i$ is the potential function of electrode $i$ ($S$ or $T$),
and $\psi^i$  depends on both time and spatial coordinates. The
stationary states are  $\psi^i = \psi^i_{\mu} e^{-iE^i_{\mu}t/\hbar}$
with the spatial wave functions and energy eigenvalues satisfying
\begin{equation}
\left(-\frac{\hbar^2}{2m}\frac{\partial^2}{\partial z^2} +
U_{i}\right)\psi^i_{\mu} = E^i_{\mu}\psi^i_{\mu},
\label{eq.bardeen.psimu}
\end{equation}

Once the distance between the two electrodes is reduced, the
time-evolution of a state $\psi$ in the system tip-sample is governed
by the Schr\"ordinger equation containing the full potential:
\begin{equation}
i\hbar\frac{\partial \psi}{\partial t}=
\left(-\frac{\hbar^2}{2m}\frac{\partial^2}{\partial z^2} +
U_S+U_T\right)\psi.
\label{eq.bardeen.AB}
\end{equation} 
The time-evolution can be treated in perturbation theory.  At $t
\rightarrow -\infty$, the tip is far from the substrate and an
electron is stationary in a state $\psi^S_{\mu}$ of the sample. We
assume that the tip is approached slowly towards the sample and
thereby the tip potential is turned on adiabatically.  The adiabatic
consideration is reasonable since  the time-scale of electrons are
femtoseconds ($\sim 10^{-15}$sec) while the time needed to move the tip
is in seconds.  Formally we describe this  adiabatic switching of the
perturbation via a time-dependent potential
\begin{equation}
U_T(t) = e^{\eta t/\hbar}U_T, \ \mathrm{and}\  \eta > 0.
\end{equation} 
 $U_T$ is a constant and $\eta$ is small and positive. At the end of
our derivation, when we consider $\eta \rightarrow 0$,  the potential
will be constant for all times.
With the presence of the combined potential, a state $\psi^S_{\mu}$
described  by Eq.~\ref{eq.bardeen.psimu} at $t = -\infty$ will not
evolve according to  Eq.~\ref{eq.bardeen.1}. Instead it has a
probability of populating the  states of electrode $T$, denoted as
$\psi^T_{\nu}$. Our goal is to measure that probability since it is
directly related to the tunneling current.

The state $\psi$ of the whole system can be expanded in a linear
combination  of the sample and tip eigenfunctions (as calculated
before the perturbation is switched on), which form an orthogonal and
complete basis set:
\begin{equation}
\psi =
a_{\mu}(t)\psi^S_{\mu}e^{-iE^S_{\mu}t/\hbar}+\sum_{\nu=1}^{\infty}
c_{\nu}(t)\psi^T_{\nu}e^{-iE^T_{\nu}t/\hbar}.
\label{eq.bardeen.trial}
\end{equation}
In our ansatz, $a_{\mu}(t)$ and $c_{\nu}(t)$  are coefficients to be
determined by Eq.~\ref{eq.bardeen.AB} with $a_{\mu}(-\infty) =1$ and
$c_{\nu}(-\infty) = 0$. We note that in our ansatz  the time evolution
coefficients $a_{\mu}(t)$ and $c_{\nu}(t)$  is due solely to the
presence of the time dependence in $U_T(t)$. As we shall see, this
separation is  convenient because the time evolution coefficients
satisfy a relatively simple differential equation.

It is important to note that each set of wave functions $\psi^S_{\mu}$
and $\psi^T_{\nu}$ originates  from different Hamiltonians. Neither of
them  is an eigenfunction of the Hamiltonian of the combined system. A
basic assumption of  Bardeen's tunneling theory is that the two sets
of wave functions are approximately orthogonal,  $\int
\psi^{T*}_{\mu}\psi^S_{\nu} d^3 r \approx 0$.  Inserting
Eq.~\ref{eq.bardeen.trial} into Eq.~\ref{eq.bardeen.AB} and after
projection on the  state $\psi^T_{\nu}$, we obtain
\begin{equation}
i\hbar \frac{dc_{\nu}(t)}{dt} = \left\langle{\psi^T_{\nu}}\right|U_T
\left|{\psi^S_{\mu}}\right\rangle
e^{-i\left(E^S_{\mu}-E^T_{\nu}+i\eta\right)t/\hbar}+
\sum_{\lambda=1}^{\infty}c_{\lambda}(t)\left\langle{\psi^T_{\nu}}\right|U_S\left|{\Psi^T_{\lambda}}\right\rangle
e^{-i\left(E^T_{\lambda}-E^T_{\nu}\right)t/\hbar}.
\label{eq.bardeen.firstorder}
\end{equation}
Here we considered the following small approximations: (i) because of
the adiabatic approximation we considered the  prefactor $a(t)$ to be
slowly varying, i.e. $\frac{d}{dt}a_{\mu}(t) = 0$ and $a_{\mu}(t) =
1$, and (ii) a second  contribution $\sim \left(e^{\eta t/\hbar}-1
\right)$ is neglected since it vanishes at ${\eta\rightarrow 0}$.

This equation can be solved iteratively but we limit ourselves to the
first order  of time-dependent perturbation theory and neglect  the
second term on the right-hand side of the previous equation since it
is a  second-order infinitesimal quantity. Therefore, to first-order, 
\begin{equation}
i\hbar \frac{dc_{\nu}(t)}{dt}
=\left\langle\psi^T_{\nu}\right|U_T\left|\psi^S_{\mu}\right\rangle
e^{-i(E^S_{\mu}-E^T_{\nu}+i\eta)t/\hbar}.
\end{equation}
Since $U_T$ is non-zero only in the volume of electrode $T$ (at $z > s$, 
see Fig.~\ref{stm_scheme}), the integral  $\left\langle
\psi^T_{\nu}\right|U_T\left|\psi^S_{\mu}\right\rangle$, that defines
the tunneling matrix element $M_{\mu\nu}$,  is evaluated only in the
right-hand side of the separation  surface. This tunneling matrix
element describes the projection of the initial state $\psi^S_{\mu}$
perturbed by the  potential $U_T$ onto the final state $\psi^T_{\nu}$
. After integration over time we get
\begin{equation}
c_{\nu}(t)= \frac{1}{E_{\mu}-E_{\nu}+i\eta}M_{\mu\nu}
e^{-i(E^S_{\mu}-E^T_{\nu}+i\eta)t/\hbar}.
\end{equation}
$|c_{\nu}(t)|^2$ describes the probability that an electron initially
described by the state $\psi^S_{\mu}$ in time $t = -\infty$ populates
a  state $\psi^T_{\nu}$ at time $t$,
\begin{equation}
|c_{\nu}(t)|^2 = \frac{e^{2\eta
    t/\hbar}}{\left(E^S_{\mu}-E^T_{\nu}\right)^2+\eta^2}\left|
M_{\mu\nu}  \right|^2,
\end{equation}
which leads to the tunneling probability per unit time, $P_{\mu\nu}(t)
= \frac{d}{dt}|c_{\nu}(t)|^2$,
\begin{equation}
P_{\mu\nu}(t) =
\frac{2\eta}{\left(E^S_{\mu}-E^T_{\nu}\right)^2+\eta^2} e^{2\eta
  t/\hbar}\frac{1}{\hbar}  \left| M_{\mu\nu}  \right|^2,
\end{equation}
where we can recognize the definition of the $\delta(x)$ distribution
given by $\delta(x) = \frac{1}{\pi}\lim\limits_{\eta \rightarrow 0}
\frac{\eta}{x^2+\eta^2}$. Taking   the limit $\eta \rightarrow 0$ we
find
\begin{equation}
P_{\mu\nu}(t) =
\frac{2\pi}{\hbar}\delta\left(E^S_{\mu}-E^T_{\nu}\right)
|M_{\mu\nu}|^2,
\end{equation}
which is the famous Fermi's Golden Rule\iffindex{Golden Rule}, 
that is a general result of
first order time-dependent perturbation theory (The Golden Rule is also derived in 
Chapter {\bf A 5} by Ph.~Mavropoulos). Elastic tunneling is
guaranteed by  the delta function
$\delta\left(E^S_{\mu}-E^T_{\nu}\right)$. The tunneling current $I$ is
proportional to  $eP_{\mu\nu}$ where $e$ is the elementary electron
charge. 

Up to now we have considered the tunneling process involving a single
state $\mu$ to a single state $\nu$. However the tip  and substrate
are characterized by  a continuous spectrum of states, thus we have to
consider the sum over states $\mu$ and $\nu$ for every
spin channel. Naturally an electron  can only tunnel from  an occupied
state $\psi^S_{\mu}$ to an unoccupied state $\psi^T_{\nu}$ and
vice-versa. At zero temperature, there is a sharp  Fermi edge
separating occupied and unoccupied states while at elevated temperatures
the  Fermi edge is smeared out; occupied states are then described by the
Fermi-Dirac distribution   $f(E-E_F) = \left(1+
exp\left[(E-E_F)/k_BT\right] \right)^{-1}$ while unoccupied states are
described by $1 - f(E-E_F)$. Accounting for the  occupation in this
manner and assuming a bias voltage $V$, the tunneling current in
thermal equilibrium from sample to tip, $I_{S\rightarrow T}$,  and
from tip to sample, $I_{T\rightarrow S}$ can be written as: 
\begin{eqnarray}
I_{{S}\rightarrow {T}} &=& \frac{4\pi e}{\hbar}\sum_{\nu\mu}
f(E^S_{\mu}-E^S_F)\left[ 1 - f(E^T_{\nu}-E^T_F)\right]\left|
M_{\mu\nu}  \right|^2 \delta(E^T_{\nu}-E^S_{\mu}-eV)\\ \nonumber
I_{{T}\rightarrow {S}} &=& \frac{4\pi e}{\hbar}\sum_{\nu\mu}
f(E^T_{\mu}-E^T_F)\left[ 1 - f(E^S_{\nu}-E^S_F)\right]\left|
M_{\mu\nu}  \right|^2 \delta(E^T_{\nu}-E^S_{\mu}-eV)
\end{eqnarray} 
A factor 2 has been introduced accounting for the two possible spin
states of each electron. The difference between the two currents gives
a net total  tunneling current:
\begin{equation}
I = \frac{4\pi e}{\hbar}\sum_{\nu\mu} \left[ f(E^S_{\mu}-E^S_F) -
  f(E^T_{\nu}-E^T_F)\right]\left| M_{\mu\nu}  \right|^2
\delta(E^T_{\nu}-E^S_{\mu}-eV).
\end{equation}
The finite summation over the discrete states can be replaced by an
integral over energies using the density of state $n(E)$:
$\sum_{\mu}\rightarrow \int n(E)dE$  and after an appropriate
change of variable we find
\begin{eqnarray}
I &=& \frac{4\pi e}{\hbar} \int d\epsilon \left[ f(E_F^T -eV +
  \epsilon) - f(E_F^S+\epsilon)\right] \\\nonumber && \times
n^T(E_F^T-eV+\epsilon)n^S(E_F^S+\epsilon) \left|
M(E_F^S+\epsilon,E_F^T-eV+\epsilon)  \right|^2
\label{eq.bardeen.M.usual}
\end{eqnarray} 
where $n^S$ and $n^T$ are the density of states (DOS) of the
substrate and of the tip. We find formally that the tunneling current
depends explicitly on  the electronic structure of both the tip and
substrate which has important consequences on STM measurements.
 Interestingly at zero temperature or if $k_BT$ is smaller
than the  energy resolution required in the measurement, the Fermi
distribution function can be approximated by a step function and the
current simplifies  to 
\begin{equation}
I =  \frac{4\pi e}{\hbar} \int_0^{eV} d\epsilon
n^T(E_F^T-eV+\epsilon)n^S(E_F^S+\epsilon) |M|^2
\end{equation} 
and for a very small bias voltage
\begin{equation}
I =  \frac{4\pi e}{\hbar} V  n^T(E_F^T)n^S(E_F^S) |M|^2.
\end{equation}

The differential conductivity, which is the other quantity measured
experimentally, is given by 
\begin{equation}
\frac{dI}{dV} =  \frac{4\pi e}{\hbar} n^T(E_F^T)
n^S(E_F^S+eV)|M(E_F^S+eV,E_F^T)|^2.
\end{equation}
This explains the unique power of STM to be able to access the occupied and unoccupied 
electronic states of the substrate. Indeed this can be achieved by changing the sign of 
the bias voltage $V$.

\subsubsection*{Inelastic tunneling \label{inelastic}}\iffindex{Inelastic Tunneling}

Although the rest of this lecture is devoted to elastic tunneling 
phenomena, inelastic tunneling within the Bardeen approach is described briefly 
in this section. 
Within STM, these excitations were observed for vibrations (see e.g.~\cite{ho1998}), 
photons (see e.g.~\cite{ho2003}) and for 
spin-excitations (see e.g.~\cite{heinrich,heinrich2,
wulfhekel,khajetoorians,Wulfhekel2008}).
\begin{figure}[ht!]
\begin{center}
\includegraphics*[width=.25\linewidth,angle=0,trim=150mm 0mm 0mm
  25mm,angle=90]{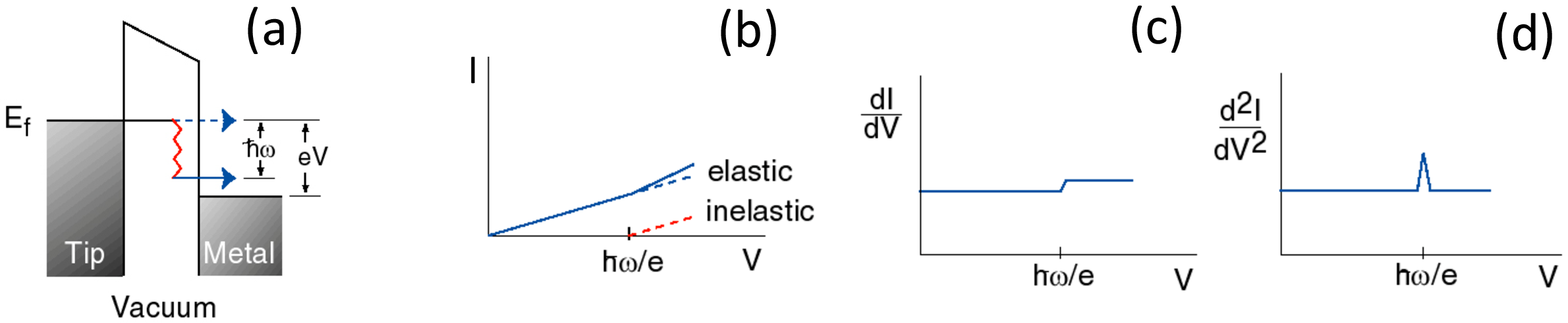}
\caption{Schematic representation of inelastic tunneling with STM (according to Wilson Ho) 
is shown in (a). 
If the electron has enough energy, provided by the bias potential, 
to trigger the excitation mode an additional tunneling channel is created. The slope 
giving the tunneling current versus the bias voltage (b) changes at a bias voltage 
corresponding to the frequency of 
the excitation mode. Taking the first and second derivatives lead to a step-like function (c) 
or to a resonance (d) at $\hbar \omega$.}
\label{inelastic_stm}
\end{center}
\end{figure}

We have learned earlier that the current and differential conductivity 
is proportional to the DOS of the substrate and of the tip. Now imagine 
that on the substrate a molecule is deposited which is characterized by 
a vibrational mode or a spin excitation mode. 
If the tunneling current can trigger the excitation, i.e. the tunneling electrons couple to the 
excitation mode and by that loose their energy, additional tunneling channels can be created. 
More tunneling possibilities translates to an increase in the tunneling current. 
Fig.~\ref{inelastic_stm} shows schematically how the slope of the current versus the bias 
voltage increases suddenly at the voltage, or energy, corresponding to the excitation mode.
If one calculates or measures the differential conductivity, one obtains step-like functions in the spectra, and a 
second derivative of the current leads to resonances located at the excitation energies.

Consider that the potential of the sample $U_S$ contains a vibrating adatom and is time-dependent
$U_S + U_0\cos(\omega t)$, 
where $U_0$ is the amplitude of the vibration and $\omega$ is the vibrational frequency of the adatom.  
We apply once more first order time-dependent perturbation theory as discussed previously  and find:
\begin{equation}
i\hbar \frac{dc_{\nu}(t)}{dt} = M_{\mu\nu}e^{-i(E^S_{\mu}-E^T_{\nu}+i\eta)t/\hbar} + 
\delta M_{\mu\nu} \cos(\omega t) e^{-i(E^S_{\mu}-E^T_{\nu})t/\hbar}
\label{eq.inelastic}
\end{equation}
where $M_{\mu\nu}$ is the elastic tunneling matrix element 
$\left\langle\psi^T_{\nu}\right|U_T\left|\psi^S_{\mu}\right\rangle$
and 
$\delta M_{\mu\nu}$ is the inelastic counterpart
$\left\langle\psi^T_{\nu}\right|U_0\left|\psi^S_{\mu}\right\rangle$.

After integration, the inelastic contribution to the coefficient $c_{\nu}$ in Eq.~\ref{eq.inelastic} is
\begin{equation}
\delta c_{\nu}(t) =\frac{\delta M_{\mu\nu}}{2} \left[ \frac{e^{(E_{\mu}-E_{\nu}+\hbar \omega)t/\hbar}}
{E_{\mu}-E_{\nu}+\hbar \omega} 
+  \frac{e^{(E_{\mu}-E_{\nu}- \hbar \omega)t/\hbar}}{E_{\mu}-E_{\nu}-\hbar \omega}  \right].
\end{equation}
The probability  is thus simply given by\footnote{Other cross terms are not considered here}
\begin{equation}
{|\delta c_{\nu}|^2} = 
 |\delta M_{\mu\nu}|^2 \left(\frac{\sin^2[(E_{\mu}-E_{\nu}+\hbar \omega)t/2\hbar]}{(E_{\mu}-E_{\nu}+\hbar \omega)^2}
+ \frac{\sin^2[(E_{\mu}-E_{\nu}-\hbar \omega)t/2\hbar]}{(E_{\mu}-E_{\nu}-\hbar \omega)^2}
+ \ \mathrm{cross\ terms}\right).\nonumber
\end{equation}
The function $\sin^2(xt/2\hbar)/x^2$ reaches its maximum when $x = 0$ and approach rapidly zero for $x \neq 0$. 
In the limit of long time  $t$, this function is 
nothing else than a delta function: $t \pi\delta(x)/2\hbar$, which ensures that the inelastic tunneling occurs at 
$E_{\mu}-E_{\nu}-\hbar =\pm \hbar \omega$. Finally we give the probability rate as done in the previous 
section:
\begin{equation}
\delta P_{\mu\nu} =\frac{d|\delta c_{\mu}|^2}{dt} \propto \frac{\pi}{2\hbar}|\delta M_{\mu\nu}|^2
\delta(E_{\mu}-E_{\nu}\pm \hbar \omega)
\end{equation}

Therefore in addition to the elastic tunneling term, we have an additional term when the excitation is created, i.e. the energy level 
of the electronic state changes by an amount $\hbar \omega$. We note that several theoretical methods were developed in order to understand how STM probes excitations. For instance, 
for spin-excitations, some are based on a Heisenberg Hamiltonian
 (see e.g. Refs.\cite{Lorente2009,Persson,Fernandez,Fransson}) and beyond\cite{Hurley},
 others are based on time-dependent density functional theory~\cite{Lounis2010,Lounis2011}.

\subsubsection*{Bardeen's Tunneling Matrix elements}
We derived earlier the tunneling current and the differential
conductivity in the Bardeen's approach, we   investigate in the following 
the tunneling matrix element $M_{\mu\nu}$.  
Using Eq.~\ref{eq.bardeen.psimu}, the integral defining the tunneling
matrix element  $M_{\mu\nu} = \left\langle
\psi^T_{\nu}\right|U_T\left|\psi^S_{\mu}\right\rangle$  can be
converted into a surface integral only depending  on the unperturbed
wave functions of the two electrodes at the separation surface. By
applying Eq.~\ref{eq.bardeen.psimu}, we have
\begin{equation}
M_{\mu\nu} =
\int_{z>z_0}\psi^S_{\mu}\left(E^T_{\nu}+\frac{\hbar^2}{2m}\frac{\partial^2}{\partial
  z^2}\right)\psi^{T*}_{\nu}d^3r.
\end{equation}
Because of the elastic tunneling condition, $E^S_{\mu} = E^T_{\nu}$,
the form giving the tunneling matrix element can be converted into
\begin{equation}
M_{\mu\nu}=\int_{z>z_0}\left(\psi^{S}_{\mu} E^S_{\mu}\psi^{T*}_{\nu}+
\psi^S_{\mu}\frac{\hbar^2}{2m}\frac{\partial^2}{\partial
  z^2}\psi_{\nu}^{T*}\right)d^3r.
\end{equation}
Using  $M_{\mu\nu} = \left\langle
\psi^S_{\mu}\right|U_T\left|\psi^T_{\nu}\right\rangle$ and  noticing
that, on the tip side, the sample potential $U_S$ is zero, we obtain
\begin{equation}
M_{\mu\nu}=
-\frac{\hbar^2}{2m}\int_{z>z_0}\left(\psi_{\nu}^{T*}\frac{\partial^2\psi_{\mu}^S}{\partial
  z^2} -   \psi_{\mu}^{*S}\frac{\partial^2\psi_{\nu}^T}{\partial
  z^2}\right)d^3 r.
\label{eq.bardeen.final.M2}
\end{equation}
With the identity
\begin{equation}
\psi_{\nu}^{T*}\frac{\partial^2\psi_{\mu}^S}{\partial z^2} -
\psi_{\mu}^{S}\frac{\partial^2\psi_{\nu}^{T*}}{\partial z^2} =
\frac{\partial}{\partial z}\left[\psi_{\nu}^{T*}\frac{\partial
    \psi_{\mu}^S}{\partial z} -  \psi_{\mu}^S\frac{\partial
    \psi_{\nu}^{T*}}{\partial z}\right],
\end{equation}
the integration over $z$ can be carried out to obtain
\begin{equation}
M_{\mu\nu} =
\frac{\hbar^2}{2m}\int_{z=z_0}\left[\psi^{S}_{\mu}\frac{\partial
    \psi_{\nu}^{T*}}{\partial z} -  \psi_{\nu}^{T*}\frac{\partial
    \psi_{\mu}^S}{\partial z}\right] dx dy.
\label{eq.bardeen.final.M}
\end{equation}

The last equation gives Bardeen's tunneling matrix element in a
one-dimensional form. It  is a surface integral of the wave functions
(and its normal derivatives) of  the two free electrodes, evaluated at
the separation surface. The potential barrier  information does not
appear explicitly, and only the information of the  wave functions at
the separation surface is required. Furthermore, the formula is
symmetric  with regards to both electrodes. It is the basis of the
reciprocity principle in STM, which has important consequences in
designing and interpreting  experimental results.

Although derived for the one-dimensional case, the Bardeen approach
can be extended to the  three-dimensional case where the tunneling matrix element, 
Eqs.~\ref{eq.bardeen.final.M2} and~\ref{eq.bardeen.final.M}, change to
\begin{equation}
M_{\mu\nu} =\frac{\hbar^2}{2m}\int_{\Omega_T}\left[\psi^S_{\mu}
  \Delta\psi_{\nu}^{T*} -  \psi_{\nu}^{T*} \Delta
  \psi_{\mu}^S\right]\cdot d\vec{r}
\label{eq.bardeen.final.M2.3D}
\end{equation}
and
 \begin{equation}
M_{\mu\nu} =\frac{\hbar^2}{2m}\int_{\Sigma}\left[\psi^S_{\mu}
  \vec{\nabla}\psi_{\nu}^{T*} -  \psi_{\nu}^{T*}
  \vec{\nabla}\psi_{\mu}^S\right]\cdot d\vec{S}
\label{eq.bardeen.final.M.3D}
\end{equation}
where the surface integral is performed on the separation surface,
$\Sigma$, between the volume defining the sample and the volume
defining the tip.

\subsubsection*{Energy dependence of tunneling matrix elements}
The assumption, that the tunneling matrix element $M$ is a constant,
is reasonable for a small bias voltage window. However, in scanning
tunneling spectroscopy (STS)  experiments, the energy scale can be as
large as $\pm 2 eV$. Thus the energy dependence of the tunneling matrix
element cannot be overlooked. The variation of  $|M|$ with energy can
be evaluated from the Bardeen formula, Eq.~\ref{eq.bardeen.final.M}.

In the gap region, the wave function of electrode $S$ is:
\begin{equation}
\psi^S_{\mu}(z) = \psi^S_{\mu}(0)e^{-\varkappa^S_{\mu}z},
\end{equation}
where $\varkappa^S_{\mu}=\sqrt{2m|E^S_{\mu}|}/\hbar$ is the decay constant
corresponding to the energy eigenvalue of $\psi^S_{\mu}$. Similarly,
in the gap region, the wave function  of electrode $T$ is
\begin{equation}
\psi^T_{\nu}(z) = \psi^T_{\nu}(s)e^{\varkappa^T_{\nu}(z-s)}.
\end{equation}
Because of the condition of elastic tunneling ($E^S_{\mu}=E^T_{\nu}$),
the two decay constants are equal,
\begin{equation}
\varkappa^T_{\nu}=\varkappa^S_{\mu} = \frac{\sqrt{2mE^S_{\mu}}}{\hbar}
\end{equation} 

Inserting the previous equations into Eq.~\ref{eq.bardeen.final.M}, we
obtain

\begin{equation}
M_{\mu\nu} =\frac{\hbar^2}{2m}e^{-\varkappa^S_{\mu}s}\int_{z=z_0} 2
\varkappa^S_{\mu}\psi^S_{\mu}(0)\psi^T_{\nu}(s) dxdy.
\end{equation}
As expected, the tunneling matrix element is independent of the
position of the separation surface, $z = z_0$.  The expression in the
integral is a constant, because $\psi^T_{\nu}(s)$ is the value of the
wave function of electrode $T$ at its surface.  The energy dependence
of $M$ is through the decay constant $\varkappa^S_{\mu}$.  Qualitatively, the
effect of the energy dependence of the tunneling matrix element on the
tunneling current is  as follows. Once the integration over energies
is carried out in Eq.~\ref{eq.bardeen.M.usual}, we realize that  the
value of $e^{-\varkappa^S_{\mu}s}$ near the top of the  integral is bigger
than its value near the bottom. Therefore, the energy spectrum of
electrode $T$ near the Fermi level and the empty states energy
from the spectrum of electrode $S$ electrons about 
eV above the Fermi level are the dominant  contributors
to the integral in Eq.~\ref{eq.bardeen.M.usual}. 

We have learned from the Bardeen's approach to calculate the tunneling
current, that the exact electronic structure  of tip and
sample is required. In principle it is possible to calculate them for
both systems with the actually available  ab-initio methods and to
compute all tunneling matrix elements to gain the tunneling
current. Although quite elaborate, this  scheme is possible. However,
the tip structure is not straightforward to access experimentally
which complicates the task of simulating the  STM tip. This issue
pushed the development of models, as the one described in the next
section, that simplify the tip's electronic structure. The simplest approach 
is to get rid off the tip.

\subsection{Tersoff-Hamann model}\iffindex{The Tersoff-Hamann model}
After the invention of STM, Tersoff and Hamann formulated a 
model~\cite{Tersoff83.1,Tersoff85.1}
based on Bardeen's tunneling theory which is widely used today. Here we
describe its concept, derivation and limitations. 
\subsubsection*{The essence of the model}
The driving argument behind the Tersoff-Hamann (TH) model is the
difficult access to the tip states. Those, as we have learned in the
previous sections, are important in the imaging mechanism of STM since
the tunneling current is a convolution of  electronic states of the
tip to those of the sample. Therefore, a particular model of the tip
was proposed, such that the tip properties can be simplified and
factorized out of the problem.  The TH model represents the tip with
potential and wave functions arbitrarily localized, in words, modeled
as a  geometrical  point. Consequently, the STM image is related to
the properties of the surface alone.  Thus, according to that model,
STM measures an intrinsic property of the unperturbed  surface, rather
than a property of the joint surface-tip system.

The TH model has proven to be extremely valuable in interpreting the
STM images with characteristic  feature sizes of $\geq 10$\AA, for
example, the  profiles of superstructures of surface reconstruction,
the scattered waves of surface states,  as well as defects,
adsorbates, and substitution atoms on the  surface. However, the TH
model predicts that the corrugation of atomic-scale features  (with
typical length scale close to or smaller than 3\AA) is about one
picometer or even smaller,  which is beyond the  detection limit of
STM. Also it cannot always explain the rich experimental observations
due, obviously, to the convolution  of tip electronic states and sample electronic
states.

\subsubsection*{Derivation of TH model}

The STM tip is modeled as a locally spherical potential well 
centered at $\vec{R}_T$. Once more, the sample
surface is represented by the $z = 0$ plane. In Bardeen's model the
potentials of the tip and sample are negligible in  the separation
plane $\Sigma$. Therefore in the vacuum region, both wave functions of
sample and tip near  the Fermi level satisfy the Schr\"odinger
equation
\begin{equation}
-\frac{\hbar^2}{2m} \Delta \psi = -\Phi\psi \ \ \mathrm{or}
\ \   \Delta \psi = \varkappa^2 \psi, 
\label{eq.tersoff.1}
\end{equation} 

With the approximation that the tip is just a single atom which has an
$s$-orbital as wave function, this  equation has two solutions, an
irregular and a regular solution,  which are the spherical, modified
Bessel functions of first and second kind.  We are interested in the
regular solution that is characterized by an exponential decay from
tip to vacuum. Thus the solution of the previous equation is
given by the modified Bessel function of  the second kind
\begin{equation}
\psi_{\nu}^T(\vec{r}-\vec{R}_T) = Ck_0^{(1)}(\varkappa|\vec{r}-\vec{R}_T|) =
C\frac{e^{-\varkappa|\vec{r}-\vec{R}_T|}}{\varkappa|\vec{r}-\vec{R}_T|}
\end{equation}  
where $|\vec{r}-\vec{R}_T|\neq 0$ since the solution is obtained in
the vacuum at position $\vec{r}$ and $C$ is a  normalization constant.

Inserting this ansatz for the tip wave function into the expression
for the matrix tunneling element (Eq.~\ref{eq.bardeen.final.M2.3D})
yields
\begin{equation}
M_{\mu\nu}(\vec{R}_T)= -
\frac{C\hbar^2}{2m}\int_{\Omega_T}\left[k_0^{(1)}(\varkappa|\vec{r}-\vec{R}_T|)
  \Delta \psi^S_{\mu}(\vec{r}-\vec{R}_T)  -\psi_{\mu}^S
  (\vec{r}-\vec{R}_T) \Delta k_0^{(1)}(\varkappa|\vec{r}-\vec{R}_T|)   \right]
d\vec{r}.
\label{eq.derivation}
\end{equation}

Since the sample potential vanishes in the body of the tip we can
apply the vacuum Schr\"odinger equation to the  first term of the
integrand. The second term has a singularity at $\vec{r}=\vec{R}$ and
can be simplified recalling the relation between  the modified Bessel
function $k_0^{(1)}$ and the Green function of the vacuum Schr\"odinger
equation:
\begin{equation}
\Delta
G(\vec{r}-\vec{r}^{\prime})=-4\pi\delta(\vec{r}-\vec{r}^{\prime}).
\end{equation} 
Since $G(\vec{r}-\vec{r}^{\prime}) = \varkappa k_0^{(1)}(\varkappa|\vec{r}-\vec{r}^{\prime}|)
$, we  rewrite $\Delta k_0^{(1)}$ in Eq.~\ref{eq.derivation} as $(\varkappa^2 k_0^{(1)} -
4\pi\delta/\varkappa)$ and by  that the tunneling matrix element for the case
of a $s$-wave function simplifies in the TH model to
\begin{equation}
M_{\mu\nu}(\vec{R}_T) = - \frac{2\pi C\hbar^2}{\varkappa m}\psi_{\mu}^S(\vec{R}_T)
\end{equation}

This is the central result of the TH model of STM although the
original derivation is a bit longer.   If the tip state is spherically
symmetric around a point  $\vec{R}_T$, effectively, it is equivalent
to a geometrical point at $\vec{R}_T$.  Hence, the tunneling matrix
element is directly  proportional to the value of the sample  wave
function at the position of the apex atom. Now we are able to
calculate the tunneling current following  Bardeen's formulation at
low temperature:
\begin{equation}
I(\vec{R}_T,V) =
\frac{16\pi^3C^2\hbar^3e}{\varkappa^2m^2}n^T\int_0^{eV}d\epsilon n^S(\vec{R}_T,E_F^S+\epsilon)
\label{eq.current.TH}
\end{equation}
where $n^T$ is a constant in the TH model since the wave function
of the tip is of $s$-type and  $n^S(\vec{R}_T,\epsilon) = \sum_{\mu}
\delta(E^S_{\mu} -
\epsilon)|\psi_{\mu}^S(\vec{R}_T)|^2$. Eq.~\ref{eq.current.TH} expresses
that the integral includes all states of the sample at the tip
location between the Fermi energy and the Fermi energy shifted  by the
applied bias voltage: the tunneling current is proportional to the
integrated local density of states (ILDOS) of the sample.

For a small bias voltage, the previous equation simplifies further to
\begin{equation}
I(\vec{R}_T,V) = \frac{16\pi^3C^2\hbar^3e^2}{\varkappa^2m^2}V n^T
n^S(\vec{R}_T,E_F^S).
\end{equation}

For the differential conductivity we obtain for finite $V$
\begin{equation}
\frac{dI(\vec{R}_T,V)}{dV} = \frac{16\pi^3C^2\hbar^3e}{\varkappa^2m^2}n^T
n^S(\vec{R}_T,E_F^S+eV).
\end{equation}

Thus the tunneling current and the differential conductivity are
proportional to the density of states $n^S(\vec{R}_T,E_F^S+eV)$  in
the vacuum. The last three equations are the most used ones in the
interpretation, simulation and prediction of STM-images for  realistic
systems. We note that an absolute value of the current within this
scheme cannot be computed since the constant $C$ is unknown.  From the
TH model, one can simulate the important STM modes mentioned in 
Section ~\ref{description}. For example, in the  constant-current mode ($I =$ const.),
where topographic images are obtained,  we use 
$I \sim eV n^S(\vec{R}_T,E_F^S)$ valid for $eV << \Phi$. Hence, the task is simply to look for
$n^S(\vec{R}_T,E_F^S) = \mathrm{const.}$ In the spectroscopic mode, $dI/dV$  is computed from 
$\sim n^S(\vec{R}_T,E_F^S+eV)$,  in other words the calculation of the
spectroscopic images obtained with STM boils down to the computation
of the  sample's DOS in vacuum.

The basic assumption of this extremely simple result  is that, except
for the $s$-wave tip wave function, all other tip wave functions can be
neglected. Therefore it is often called the s-wave  tip model. It is
important to know under which condition the s-wave-only assumption is
valid. 
The TH model is highly valuable in the interpretation of STM
images. It  represents an approximation with which the complicated
problem of tip  electronic states can be avoided.

What if the STM-tip is more complicated than what is assumed in the 
TH-model? For example, other orbitals, than the $s$, can 
be characterizing the apex atom. Chen~\cite{Chen90.1,Chen90.2} proposed an elegant 
method, discussed in the upcoming subsection, that extends the TH-model.

\subsubsection*{Chen's expansion of the Tersoff-Hamann model}\iffindex{Chen's expansion of the Tersoff-Hamann model}
The problems which arise from the Tersoff-Hamann model can be overcome
by  expanding the model using generalized wave functions for the
tip. Such an expansion was introduced by  Chen~\cite{Chen90.1,Chen90.2} 
who considered
the general solutions of the vacuum Schr\"odinger equation
(Eq.~\ref{eq.tersoff.1})
\begin{equation}
\psi^T(\vec{r}-\vec{R}_T)=\sum_{l,m}C_{lm}k_l^{(1)}(\varkappa|\vec{r}-\vec{R}_T|)
Y_{lm}(|\widehat{\vec{r}-\vec{R}_T}|)
\end{equation}
where $Y_{lm}$ are the spherical harmonics and $k_1^{(1)}$ are modified
spherical Bessel functions of the  second kind while $C_{lm}$ is a
renormalization coefficient. For the case $l = 0$ we recover the TH
model as detailed in the previous  subsection. Using the property of
the Bessel function:
\begin{equation}
k_l^{(1)}(u) = (-1)^lu^l\left(\frac{1}{u}\frac{d}{du} \right)^lk_0^{(1)}(u)
\end{equation}
one can evaluate the contribution of a tip-orbital $l$ to the
tunneling current just by proceeding to the $l^{th}$ derivative  with
respect to the argument of modified Bessel function with $l = 0$. For
example, if the tip is described by a $p_z$-type  orbital we have
\begin{eqnarray}
\psi_{p_z}^T(\vec(r)-\vec{R}_T) &=&
C_{p_z}k_1^{(1)}(\varkappa|\vec{r}-\vec{R}|)Y_{10}(|\widehat{\vec{r}-\vec{R}_T}|)\\\nonumber
&=&C_{p_z}\frac{d}{dR}k_0^{(1)}(\varkappa|\vec{r}-\vec{R}_T|)\frac{\partial
  R}{\partial z}\\\nonumber
&=&\frac{C_{p_z}}{C_s}\frac{\partial}{\partial z}\psi^T_s(\vec{r}-\vec{R}_T).
\end{eqnarray}
 Inserting this wave
function in the Bardeen's formula for the tunneling matrix elements
in the TH model,  we obtain for the $p_z$ orbital
\begin{equation}
M_{\mu\nu}=-\frac{2\pi C_{p_z}\hbar^2}{\varkappa m}\frac{\partial}{\partial
  z}\psi_{\mu}^S(\vec{R}_T).
\end{equation}
This means, that the matrix element is proportional to the derivative
of the sample wave function with respect to $Z$ at the position of the tip, if the tip is
described by a $p_z$-type orbital. In this way the matrix element can
be derived also for higher order orbitals, which is known as the
derivative rule of Chen~\cite{Chen90.1,Chen90.2}. 

Using $p$- or $d$-type  orbitals for the tip, the experimentally
observed corrugation amplitudes of densely packed metal surfaces can
be explained. In  Table~\ref{M.chen} the tunneling matrix
elements are given for different orbitals. With the help of this rule, Chen
could explain the  high corrugation amplitude observed in some
systems.

\begin{table}
\begin{center}
\begin{tabular}{cc}
\hline & \\ Orbital of the tip & Matrix element M \\ & \\ \hline &
\\ $s$      &   $\frac{2\pi C\hbar^2}{\varkappa m} \psi^S(\vec{R}_T)$\\ &
\\ $p_x$    &   $\frac{2\pi C\hbar^2}{\varkappa m} \frac{\partial}{\partial x}
\psi^S(\vec{R}_T)$ \\ & \\ $p_y$    &   $\frac{2\pi C\hbar^2}{\varkappa m}
\frac{\partial}{\partial y} \psi^S(\vec{R}_T)$ \\  & \\ $p_z$    &
$\frac{2\pi C\hbar^2}{\varkappa m} \frac{\partial}{\partial z} \psi^S(\vec{R}_T)$
\\ & \\ $d_{zx}$  &  $ \frac{2\pi C\hbar^2}{\varkappa m}
\frac{\partial^2}{\partial z\partial x } \psi^S(\vec{R}_T)$ \\ &
\\ $d_{zy}$  &   $\frac{2\pi C\hbar^2}{\varkappa m} \frac{\partial^2}{\partial
  z\partial y} \psi^S(\vec{R}_T)$  \\  & \\ $d_{xy}$  &   $\frac{2\pi
  C\hbar^2}{\varkappa m} \frac{\partial^2}{\partial x\partial y}
\psi^S(\vec{R}_T)$  \\ & \\ $d_{z^2-\frac{1}{3}r^2}$ & $\frac{2\pi
  C\hbar^2}{\varkappa m} \left( \frac{\partial^2}{\partial z^2} \psi^S(\vec{R}_T)
-\frac{1}{3}\varkappa^2 \psi^S(\vec{R}_T) \right)$\\ & \\ $d_{x^2-y^2}$ &
$\frac{2\pi C\hbar^2}{\varkappa m} \left( \frac{\partial^2}{\partial x^2}
\psi^S(\vec{R}_T) -  \frac{\partial^2}{\partial y^2}\psi^S(\vec{R}_T)
\right)$\\ & \\ \hline
\end{tabular}
\caption{Tunneling matrix elements as formulated by Chen~\cite{Chen93} using
  the derivative rule.  Note that the constant $C$ depends on the 
  orbital-type of the tip involved  in the tunneling process.}
\end{center}
\label{M.chen}
\end{table}

The extension of Chen provides an explanation of the high corrugation
amplitudes measured on  close-packed metal surfaces contradictory to
the low corrugation amplitudes due to their local density of states.
This is the case, for example, for $p_z$ and 
$d_{z^2}$ orbitals since they possess charge density  stretching out further from
the tip apex into the vacuum than that of an $s$-wave and they act similar
to an  $s$-wave at a reduced distance from the sample surface. This
affects the tunneling current quite strongly and by that the images
obtained experimentally. Interestingly, orbitals like $d_{xy}$ and
$d_{xz,yz}$ are expected to produce  a large tunneling current not
with the tip apex atom located on top  of a surface atom but rather at
a hollow site of the surface. Due to their particular charge  density
distribution a large overlap with sample orbitals occurs in this
configuration. 
\subsection{Different STM modes}
Now that we know how to compute the tunneling current, it is
interesting to connect this quantity to the different  STM standard
modes. For a given bias, the physical quantity measured by the STM is
the tunneling current, which is  a function of the lateral coordinates
$(x, y)$ and the $z$-coordinate: $I = I(x,y,z)$. If $z$ is
perpendicular to  a nearly flat surface, the tunneling current can be
decomposed into a constant (that is independent of $(x, y)$)  and a
small  variable component that represents the features or corrugation
of the surface,
\begin{equation}
I(x,y,z) = I_0(z) + \Delta I(x,y,z),
\end{equation}
with the condition $|\Delta I(x,y,z)| <<  |I_0(z)|$. 

The constant-current topographic image can be derived from the current
images by making the ansatz: $z(x,y) = z_0 + \Delta z(x,y)$
and substituting it into the previous equation by proceeding to a
Taylor expansion we find
\begin{equation}
I = I_0(z_0) + \left( \frac{dI_0(z)}{dz}\right)\Delta z(x,y)+\Delta
I(z,y,z).
\end{equation}
Owing to the smallness of $\Delta I$, its variation can be
neglected. The topographic image is therefore defined by the condition
of constant current, i.e. $I = I_0(z_0)$, thus
\begin{equation}
\Delta z(x,y) = -\frac{\Delta I(x,y)}{dI_0(z)/dz}.
\end{equation}  

It is interesting to note that experimentally  when scanning the
sample surface with the tip there are two different modes of
operation, the constant-height and  the constant-current mode. In
constant-height mode, the vertical position $z$ of the tip is held
constant while  scanning and the resulting tunnel current between tip
and sample is measured.  In constant-current mode  a feedback loop
provides a constant tunnel current between tip and sample at every
position $(x, y)$.  This means that the $z$-position of the tip has to
be adjusted during scanning which is done by applying an  appropriate
voltage $V_z$ to the $z$-piezo of the tube scanner. One distinguishes
between these two extreme modes of operation even though neither of
them can be realized experimentally and one can only approximate one
or the other  by choosing the appropriate parameters for the feedback
loop gain and the scan speed.

The differential tunneling conductivity, $dI/dV$, is also a frequently
used image parameter. Experimentally the tunneling  spectrum at each
point $(x,y)$ can be obtained by interrupting the feedback circuit,
applying a voltage ramp, then acquiring the tunneling  current. The
differential tunneling conductivity can be obtained by numerically
differentiating the acquired tunneling current  data. 


\subsection{Spin-polarization and tunneling: SP-STM and TAMR-STM }
\iffindex{SP-STM} \iffindex{TAMR} If the tunnel current flows between two magnetic electrodes, an
additional  information will be contained in the tunneling current,
namely  the information on the magnetic properties of the electrodes.
Thus, if the STM tip is spin-polarized, in other words, the DOS for
the  majority-spin ($\uparrow$) channel is  different from the DOS for
the minority-spin ($\downarrow$) channel,  access to the local
spin-polarization of the probed substrate is possible 
(Figs.~\ref{compilations_sp_stm_tamr}(a) and (b)).  This is the
concept of spin-polarized STM (SP-STM).  By assuming that the electron
spin is conserved during the tunneling process,  the
$\uparrow$-electrons  from the tip can only tunnel into unoccupied
$\uparrow$-states in the sample;  the same for the
$\downarrow$-electrons (Fig.~\ref{compilations_sp_stm_tamr}(b)).  
When the magnetization directions of the two
electrodes are in parallel alignment  the tunnel current is different
compared to the antiparallel alignment.

\begin{figure}[ht!]
\begin{center}
\includegraphics*[width=.7\linewidth,trim=0mm 0mm 0mm
  0mm,angle=90]{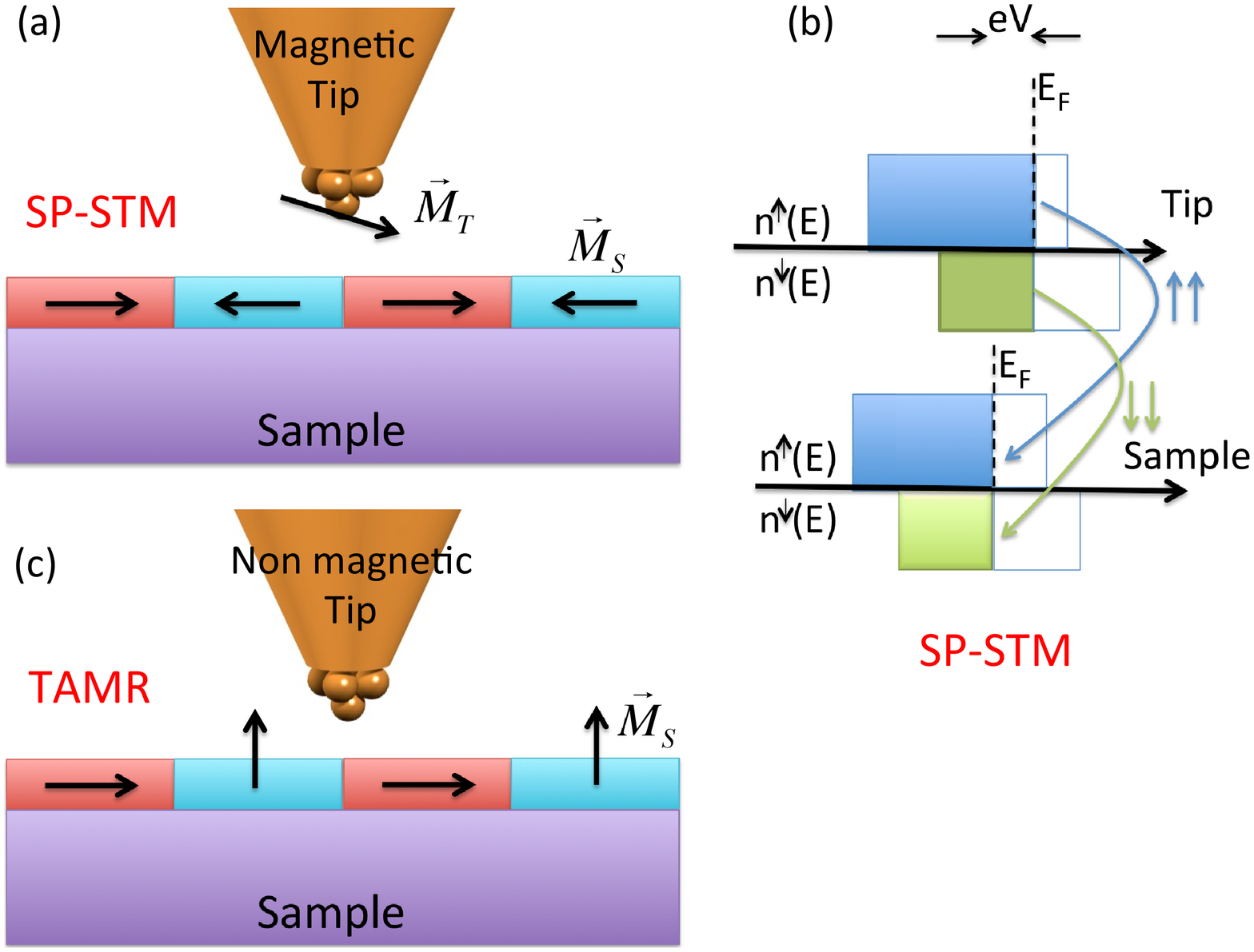}
\caption{Spin-polarized tunneling with SP-STM shown schematically in (a) and (b). In (b), the 
spin-polarized DOS of the tip and sample are depicted. With spin conservation, electrons from the tip 
with spin $\uparrow$ can tunnel into unoccupied states of the sample with the same spin character. The same 
tunneling process occurs with spin $\downarrow$. In (c) a sketch of TAMR is shown. The tunnel junction comprises 
a non-magnetic tip and a magnetic sample. Because of spin-orbit coupling the tunneling current can be sensitive to 
the magnetization's orientation of the sample.}
\label{compilations_sp_stm_tamr}
\end{center}
\end{figure}

In fact Julli\`ere~\cite{Julliere} first discovered this spin valve effect, or
tunneling  magnetic resistance (TMR), in planar Fe-Ge-Co  tunnel
junctions effect which showed a decreased conductance in the case of
non-parallel alignment of the  electrodes’ magnetization compared to
the parallel case. In a theoretical work, Slonczewski  extended the
model of tunneling in one dimension considering spin-polarized
electrodes~\cite{Slonczewski}. While the experiments  of Julli\`ere had to be
realized at very low temperatures, the TMR effect at room  temperature
was achieved in the 90's by Moodera~\cite{Moodera} and 
Miyazaki~\cite{Miyazaki}. This
allows  for the application of the TMR effect in read heads of modern
hard disk drives.  Furthermore, the discovery of the TMR gave rise to
the development of  the magneto-resistive random-access memory (MRAM)
-- a non-volatile random-access memory  technology.

Due to spin-orbit coupling
(SOC), the resistance can become anisotropic,  i.e., it depends on the
magnetization direction of the tunnel junction with respect to  the
crystallographic axes as sketched in Fig.~\ref{compilations_sp_stm_tamr}(c). 
For the observation
of this  effect the tunneling junction needs only a single magnetic
electrode separated from a nonmagnetic  electrode by an insulating
layer. In fact depending on the magnetization direction, the
electronic  structure of the magnetic electrode changes, thus the
tunneling current between the two  electrodes exhibits differences for
a film that is magnetized either out-of-plane or 
in-plane~\cite{Bode2002,Molenkamp}.
This effect has been named tunneling anisotropic magneto-resistance
(TAMR)~\cite{Molenkamp} which is an extension  of the known bulk AMR that does
not involve tunneling.  Besides its implications in spintronics, it is
appealing to use the TAMR concept in STM since even without  a
spin-polarized tip, magnetic information can be grasped from the
tunneling current if the sample  is characterized by a
non-negligible SOC. The TAMR has first been observed in STM
measurements of a double-layer film of Fe on the W(110) surface~\cite{Bode2002} 
and was recently applied on  adatoms deposited on magnetic  
substrates~\cite{Neel,Serrate}.

Since more than 10 years, SP-STM is a well-established technique which
can be used to  investigate the magnetic  ground state of
nanostructures down to the atomic scale~\cite{Heinze2000}. In the past years,
the technique has been  extended to study also dynamics of magnetic
systems like spin-flip 
processes~\cite{heinrich,heinrich2,wulfhekel,khajetoorians} 
or  magnon excitation~\cite{Wulfhekel2008}.
    Furthermore, very recently SP-STM has  been used to
probe spin relaxations of single atoms on the time scale of
nanoseconds~\cite{Loth,Khajetoorians2013} to few 
minutes~\cite{Wulfhekel2013}.

In the following two subsections, the basic theoretical concepts
behind SP-STM and TAMR within  STM  are presented.

\subsubsection*{Bardeen's formalism for SP-STM}
The Bardeen tunneling theory can be extended to include the spin 
dependence. Instead
of using a  single-component wave function, two components, i.e. a
spinor, are necessary to  describe a state of an electron with
spin. For example, for the  sample wave function
\begin{equation}
\psi_{\mu\sigma}^S = \sum_{\sigma'=\uparrow,\downarrow}^{}\psi^s_{\mu\sigma\sigma'}\chi_{\sigma'},
\ \ \mathrm{with} \ \ \chi_{\uparrow}=\left(
\begin{array}{c}
1\\ 0\end{array}
\right), \chi_{\downarrow}=\left(
\begin{array}{c}
1\\ 0\end{array}
\right).
\end{equation}

One can follow the same procedure as done previously in
Section ~\ref{one-dim} for the non spin-polarized case by considering
the time-dependent Pauli equation of the combined system 
\begin{equation}
i\hbar \frac{\partial \psi}{\partial
  t}=\left[-\frac{\hbar^2}{2m}\nabla^2+\hat{U}_T+\hat{U}_S
  \right]\psi.
\end{equation}

Note that the wave function $\psi$ is now a two-component spinor and
the potential functions are now two-by-two matrices,
\begin{equation}
\hat{U}_T = \left(
\begin{array}{cc}
U_{T\uparrow\uparrow} &
U_{T\uparrow\downarrow}\\ U_{T\downarrow\uparrow} &
U_{T\downarrow\downarrow}\\
\end{array}
\right),    \hat{U}_S = \left(
\begin{array}{cc}
U_{S\uparrow\uparrow} &
U_{S\uparrow\downarrow}\\ U_{S\downarrow\uparrow} &
U_{S\downarrow\downarrow}\\
\end{array}
\right),
\end{equation}

We follow the treatment of Wortmann {\it et al}.~\cite{Wortmann} to take the
spin-polarization direction of one of the electrodes, say, electrode
T, as  the reference (global spin frame of reference). In other words, we assume that the Hamiltonian
of the tip is diagonal with respect to spin,
\begin{equation}
\hat{U}_T = \left(
\begin{array}{cc}
U_{T\uparrow\uparrow} & 0\\ 0 & U_{T\downarrow\downarrow}\\
\end{array}
\right).
\end{equation}
Therefore, the two components of the wave functions of the tip-only system can be
treated separately to satisfy the following equation (considering the tip only),
\begin{equation}
\left[-\frac{\hbar^2}{2m}\nabla^2 + U_{T\sigma\sigma}
  \right]\psi^T_{\nu\sigma\sigma}(\vec{r}) =
E_{\nu\sigma\sigma}\psi^T_{\nu\sigma\sigma}(\vec{r}),
\end{equation}
where $\sigma$ denotes the spin component $\uparrow$ or $\downarrow$.

However, in the reference frame of the tip, the state of electrode S
is, in general, not diagonalized with respect to spin. This is
evidently true for non-collinear systems since no quantization axis
exists which allows a state to be written in terms of pure spin-up  or
spin-down character, but even for collinear samples the states will be
spin mixed if the quantization axis of the sample and  the one of the
tip are not aligned in parallel. In general, the spinor of the sample-only system satisfies the Pauli equation,
\begin{equation}
\left[-\frac{\hbar^2}{2m}\nabla^2 + \left(
\begin{array}{cc}
U_{S\uparrow\uparrow} &
U_{S\uparrow\downarrow}\\ U_{S\downarrow\uparrow} &
U_{S\downarrow\downarrow}\\
\end{array}
\right) \right] \left(
\begin{array}{c}
\psi^S_{\mu\uparrow\sigma}\\ \psi^S_{\mu\downarrow\sigma}\\
\end{array}
\right)= \hat{E}_{\mu} \left(
\begin{array}{c}
\psi^S_{\mu\uparrow\sigma}\\ \psi^S_{\mu\downarrow\sigma}\\
\end{array}
\right).
\end{equation}

 It is useful to express the
components of spinor describing the sample's wave  function in the
spin-frame of reference of the tip within the local spin frame of
reference related to the sample.  In this local spin-frame of
reference  $\hat{U}_S$ is diagonal in spin-space. This can be achieved
using the  rotation matrix $\hat{U}$.
\begin{equation}
|\psi_{\sigma}^S\rangle=\hat{U}(\theta)|\psi_{\sigma}^{S,loc}\rangle,\ \ \mathrm{with}
\ \  \hat{U}(\theta)= \left(
\begin{array}{cc}
\cos{(\theta/2)}&-\sin{(\theta/2)}\\ \sin{(\theta/2)}&
\cos{(\theta/2)}\\
\end{array}
\right),
\end{equation}
where $loc$ stands for local spin-frame of reference of the sample and
$\theta$ defines the  angle between the magnetization of the
individual atom on the sample and the magnetization of the tip.

Following the same procedure as done in the non spin-polarized case,
we find in first-order perturbation  theory that  the spin-dependent
tunneling matrix elements are given by
\begin{eqnarray}
M_{\mu\nu}^{\sigma{\sigma^{\prime}}}  & = &  \left\langle
\psi^T_{\mu{\sigma^{\prime}}}\left|\hat{U}_{T}\hat{U}(\theta)\right|\psi^{S,loc}_{\mu\sigma}\right\rangle
\end{eqnarray}
and 
\begin{eqnarray}
\left(
\begin{array}{cc}
M^{\uparrow\uparrow}_{\mu\nu} &
M^{\uparrow\downarrow}_{\mu\nu}\\ M^{\downarrow\uparrow}_{\mu\nu} &
M^{\downarrow\downarrow}_{\mu\nu}
\end{array}
\right) & = &  \left(
\begin{array}{cc}
\langle
\psi^T_{\mu{\uparrow}}|U_{T\uparrow\uparrow}|\psi^{S,loc}_{\mu\uparrow}\rangle
\cos{(\theta/2)} &-\langle
\psi^T_{\mu{\uparrow}}|U_{T\uparrow\uparrow}|\psi^{S,loc}_{\mu\downarrow}\rangle\sin{(\theta/2)}\\ \langle
\psi^T_{\mu{\downarrow}}|U_{T\downarrow\downarrow}|\psi^{S,loc}_{\mu\uparrow}\rangle
\sin{(\theta/2)} & \langle
\psi^T_{\mu{\downarrow}}|U_{T\downarrow\downarrow}|\psi^{S,loc}_{\mu\downarrow}\rangle
\cos{(\theta/2)}
\end{array}
\right)
\end{eqnarray}
\begin{eqnarray}
\ \ \ \hspace{2.5cm} &=&  -\frac{2\pi \hbar^2}{m} \left(
\begin{array}{cc}
\frac{C_{\uparrow}}{\varkappa_{\uparrow}}\psi_{\mu\uparrow}^{S,loc}(\vec{R}_T)
\cos{(\theta/2)}
&-\frac{C_{\uparrow}}{\varkappa_{\uparrow}}\psi_{\mu\downarrow}^{S,loc}(\vec{R}_T)\sin{(\theta/2)}\\ 
\frac{C_{\downarrow}}{\varkappa_{\downarrow}}\psi_{\mu\uparrow}^{S,loc}(\vec{R}_T)\sin{(\theta/2)}
&\frac{C_{\downarrow}}{\varkappa_{\downarrow}}\psi_{\mu\downarrow}^{S,loc}(\vec{R}_T)\cos{(\theta/2)}
\end{array}
\right)
\end{eqnarray}
In the last equation we followed the TH model to extract the tunneling
matrix elements by replacing   the wave function at the tip apex atom
by a spherically symmetric $s$-wave. Also the Chen's rule for
arbitrary orbitals can be followed  in the spin-polarized case. In the
following we assume that the spin-up and spin-down s-wave states can
be characterized  by the same decay constant $\varkappa$ and the same
normalization coefficient $C$, i.e. $\varkappa_{\sigma} = \varkappa$ and $C_{\sigma} =
C$.

The tunneling current in the spin-polarized case becomes
\begin{eqnarray}
I(\theta,V) &=& \frac{2\pi e}{\hbar} \sum_{\sigma\sigma^{\prime}}
\int_0^{eV} d\epsilon  \left[ f(E_F^T -eV + \epsilon) -
  f(E_F^S+\epsilon)  \right] \\\nonumber && \times
n^T_{\sigma^{\prime}}(E_F^T-eV+\epsilon)n^S_{\sigma}(E_F^S+\epsilon)
\left| M_{\sigma\sigma{\prime}}
(\theta,E_F^S+\epsilon,E_F^T-eV+\epsilon)  \right|^2
\end{eqnarray}
In the TH model, the tunneling current simplifies to
\begin{eqnarray}
I(\vec{R}_T,\theta,V) &=& \frac{16\pi^3 C^2 \hbar^3 e}{\varkappa^2m^2}
\int_0^{eV} d\epsilon  \left[ f(E_F^T -eV + \epsilon) -
  f(E_F^S+\epsilon)  \right] \\\nonumber && \times \left(
n^T(E_F^T-eV+\epsilon)n^S(\vec{R}_T,E_F^S+\epsilon) +
\vec{m}^T(E_F^T-eV+\epsilon)\cdot\vec{m}^S(\vec{R}_T,E_F^S+\epsilon)
\right)
\end{eqnarray}
where $n$ is the total charge density of states
($n=n_{\uparrow}+n_{\downarrow}$) of the tip or sample while
$\vec{m}$ is the magnetization vector, i.e. the corresponding 
"magnetic" density of states.  A further approximations can
be made by considering the tip to be characterized by an $s$-wave
function. That allows to consider  $n^T$ to be a constant:
\begin{eqnarray}
I(\vec{R}_T,\theta,V) &=& \frac{16\pi^3 C^2 \hbar^3 e}{\varkappa^2m^2}
(\underbrace{n^TN^S(\vec{R}_T,V)}_{\mathrm{non spin-polarized}} +
\underbrace{\vec{m}^T\cdot\vec{M}^S(\vec{R}_T,V)}_{\mathrm{spin-polarized}}
),
\end{eqnarray}
where $N^s$ and $M^s$ are respectively the energy integrated charge-density and
magnetization of the sample's atom at the tip location $\vec{R}_T$.
The previous equation, widely  used to interpret SP-STM, shows
that the tunneling current can be decomposed into a non spin-polarized
and spin-polarized contributions. In case of a non spin-polarized STM
experiment, i.e., using  either a nonmagnetic tip or sample, the
second term vanishes and the current reduces to the classical result
of the TH model.  Furthermore, depending on the angle between the
magnetization vectors of  tip and sample, the current will change.  

For completeness we give the corresponding differential conductivity
in the spin-polarized case
\begin{eqnarray}
\frac{dI}{dV}(\vec{R}_T,\theta,V) \propto
(\underbrace{n^Tn^S(\vec{R}_T,E_F+eV)}_{\mathrm{non spin-polarized}} +
\underbrace{\vec{m}^T\cdot\vec{m}^S(\vec{R}_T,E_F+eV)}_{\mathrm{spin-polarized}}
).
\end{eqnarray}

\subsubsection*{Tunneling anisotropic magneto-resistance (TAMR)\label{TAMR}}

We derived previously the tunneling current and the
differential conductivity in case  of a spin-polarized tip. If the tip
is non spin-polarized, the current depends solely on the non
spin-polarized part. Assume that the sample magnetization is oriented along
the out-of-plane direction,  and that one applies a magnetic field to
reorient the magnetization to be in-plane. In some cases, for
instance if SOC is present, this  reorientation could affect the
electronic structure~\cite{Lessard}. In other words, this means that 
$n^S$ then exhibits  a
dependence on the magnetization's orientation.

This can be noticed by considering the SOC potential 
\begin{equation}
\hat{V}_{SOC}  = \zeta \vec{L}\cdot\vec{\hat{S}} = \zeta \left(
\begin{array}{cc}
V_{SOC}^{\uparrow\uparrow} & V_{SOC}^{\uparrow\downarrow}
\\ V_{SOC}^{\downarrow\uparrow}  & V_{SOC}^{\downarrow\downarrow} 
\end{array}
\right)
\end{equation}
where $\zeta$ is the SOC strength, $\vec{L}$ and $\vec{S}$ are the
orbital and angular momenta. The Hamiltonian without  SOC, $H^0$, is
spin-diagonal and the unperturbed Bloch eigenfunctions are
$(\psi^{(0)}_{\vec{k}\mu\uparrow},0)^T$  and
$(0,\psi^{(0)}_{\vec{k}\mu\downarrow})^T$. The Schr\"odinger equation for the
perturbed wave function then reads
\begin{equation}
\left(
\begin{array}{cc}
H^{0\uparrow}+V_{SOC}^{\uparrow\uparrow} - E &
V_{SOC}^{\uparrow\downarrow}\\ V_{SOC}^{\downarrow\uparrow} &
H^{0\downarrow}+V_{SOC}^{\downarrow\downarrow} - E
\end{array}
\right) \left(
\begin{array}{c}
\psi^{}_{\vec{k}\mu\sigma\uparrow}\\ \psi^{}_{\vec{k}\mu\sigma\downarrow}
\end{array}
\right) = 0.
\end{equation}
The potential terms $V_{SOC}{\uparrow\downarrow}$ and
$V_{SOC}{\downarrow\uparrow}$ are  responsible for flipping the
spin. Solving this equation for the minority-spin channel, i.e. 
$\sigma = \downarrow$, leads for the two components of the spinor in first-order perturbation theory:
\begin{equation}
\psi^{(1)}_{\vec{k}\mu\downarrow\uparrow} = 
  \sum_{\nu} \frac{\langle
  \psi_{\vec{k}\nu\downarrow}^{(0)} |
  V_{SOC}^{\downarrow\uparrow}|\psi^{(0)}_{\vec{k}\mu\uparrow}\rangle}
    {E_{\vec{k}\mu}^{0\uparrow}-E_{\vec{k}\nu}^{0\downarrow}}
    \psi_{\vec{k}\nu\downarrow}^{(0)}
\end{equation}
and
\begin{equation}
\psi^{(1)}_{\vec{k}\mu\downarrow\downarrow} = \psi^{(0)}_{\vec{k}\mu\downarrow}+
  \sum_{\nu\neq\mu} \frac{\langle
  \psi_{\vec{k}\nu\downarrow}^{(0)} |
  V_{SOC}^{\downarrow\downarrow}|\psi^{(0)}_{\vec{k}\mu\downarrow}\rangle}
    {E_{\vec{k}\mu}^{0\downarrow}-E_{\vec{k}\nu}^{0\downarrow}}
    \psi_{\vec{k}\nu\downarrow}^{(0)}
\end{equation}

where the index (1) stands for the
first-order  solution and  $E_{\vec{k}\mu}^{0\uparrow}$ 
the eigenenergy of the state
$\psi_{\vec{k}\mu\uparrow}^{0}$. Since the DOS  is related to $|\psi^{(1)}|^2$, 
 it is expected that the DOS will change
because of SOC in  a quadratic fashion: $\sim
\sum\limits_{\sigma\sigma'}(V_{SOC}^{\sigma\sigma'})^2$ but the denominator of the 
previous equations will play an important role. Indeed if the weight of the states 
close to those $\vec{k}$ where the unperturbed spin-dependent bands cross, i.e. 
$E_{\vec{k}\mu}^{0\sigma} =
E_{\vec{k}\mu}^{0\sigma'}$ the denominator becomes  small and the
bands strongly couple and the modification of the charge can be important. 

To illustrate the dependence of the DOS on the SOC and rotation angle of the magnetization, 
we  introduce a simple  toy model initially proposed by N\'eel 
   {\it et al.}~\cite{Neel,Schroeder} that shows how the total  DOS, and by that the
 tunneling current and differential conductivity, can be affected by
 the magnetization orientation. 

Assume that in the energy window of interest, we have two states, say
$d_{z^2}$ and $d_{zx}$ states, in  the minority spin-channel
originally located at energies $\epsilon_1$ and $\epsilon_2$.  For
simplification, we consider full  spin-polarization, meaning that no
orbitals are present  in the majority spin channel. The hybridization
with the background is described via  $\Gamma_1$ and $\Gamma_2$. The
interaction between the two terms is provided via a hopping term $t$
which  in our case is created by SOC. The form of $t$ is inferred from
the work of Abate and Asdente~\cite{Abate} on bulk Fe  who used a
tight-binding formalism to evaluate the matrix elements $|\langle
d_m|V_{SOC}|d_{m'}\rangle|$.

The matrix element connecting
our orbitals are given by
\begin{equation}
|\langle d_{zx}|V_{SOC}|d_{z^2}\rangle| =
\frac{1}{2}\sqrt{3}\zeta\sin{\theta}\sin{\phi}
\end{equation}
for the minority spin-channel. $\theta$ and $\phi$ are the Euler angles
defining the orientation  of the magnetization.
\begin{figure}[ht!]
\begin{center}
\includegraphics*[width=.7\linewidth,angle=270,trim=20mm 0mm 0mm 0mm]{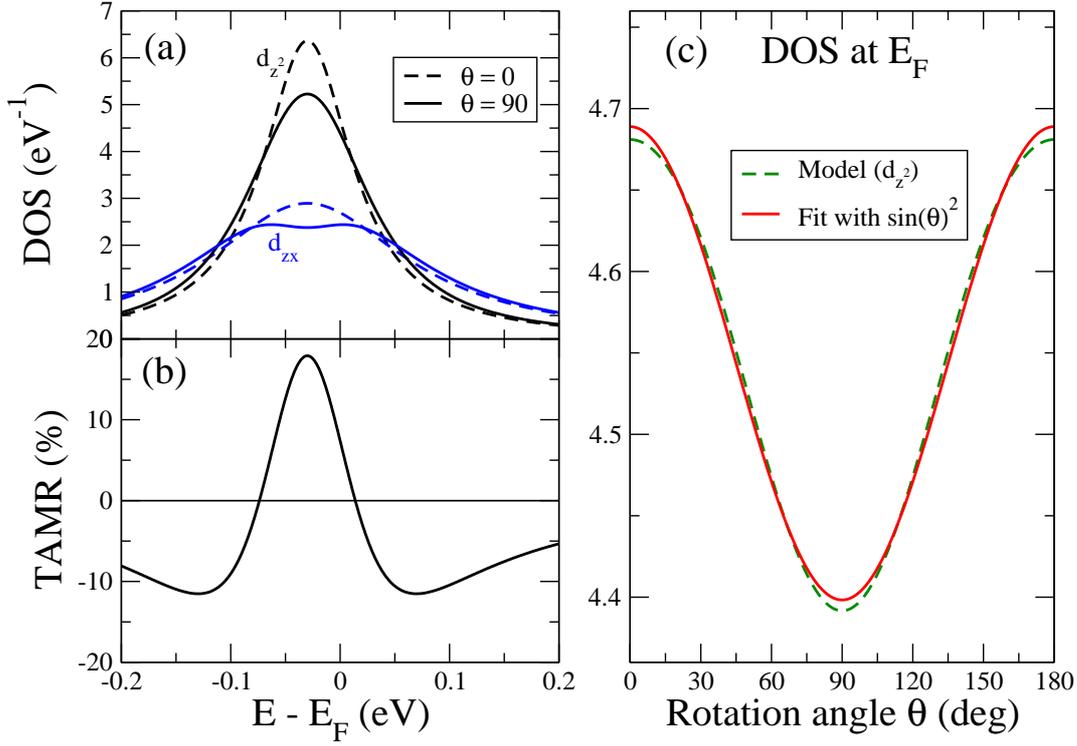}
\caption{(a) DOS obtained from a simple model for two $d$ orbitals
  interaction via SOC (see Refs.\cite{Neel,Schroeder}). Dashed and  full lines refer to the DOS
  calculated when the magnetization is out-of-plane and in-plane. The
  corresponding TAMR signal is plotted in (b) while the variation of
  the $d_{z^2}$-DOS at $E_F$ with  respect to the rotation angle
  $\theta$ is shown in (c).}
\label{fig.TAMR}
\end{center}
\end{figure}

We calculate the density state of this two-orbitals model by evaluating the Green function $G = (1 - H)^{-1}$ from which 
the imaginary part is extracted ($n(E) = -\frac{1}{\pi} \sum\limits_{m=1}^2\Im G_{mm}(E)$). We find 
\begin{equation}
\hat{G} =
\frac{1}{(E-\epsilon_1-i\Gamma_1)(E-\epsilon_2-i\Gamma_2)-t^2} \left(
\begin{array}{cc}
E-\epsilon_1-i\Gamma_1 & t\\ t & E -\epsilon_2-i\Gamma_2
\end{array}
\right),
\end{equation}
where one notices that the diagonal elements of the Green functions, $G_{mm}$ are proportional to 
$t^2$, i.e., to $V_{soc}^2$ which bears the angular dependence. Thus this toy model indicates, 
that in the presence of SOC, the local
DOS will depend on the  orientation angle of the magnetization
vector. The results obtained with  $\zeta$ =50 meV (approximate value for 3$d$ transition
elements) are
depicted in Fig.~\ref{fig.TAMR}(a) where the corresponding DOS is shown
for  two different orientations, out-of-plane ($\theta = 0^o$) and
in-plane ($\theta = 90^o, \phi = 0^o$).  One notices that changes occur for both
orbitals upon rotation. From our different formulas giving  the
tunneling current it is obvious that since the charge of the sample
gets modified,  the current magnitude will be affected by the rotation
of the magnetization vector.  In Fig.~\ref{fig.TAMR}(b) is presented
the corresponding TAMR signal, given by
$\frac{n(\theta=0)-n(\theta=90)}{n(\theta=0)}$. A value of 20\% is
found at the position of  the $d$-resonances where the large change in the
DOS is observed. Experimentally, it is thus worthwhile  to probe the
sample at those energies where the largest TAMR effect is
observed.  It is interesting to check the angle dependence of the DOS
as shown in Fig.~\ref{fig.TAMR}(c).  Interestingly, the $d_z^2$
contribution to the DOS, thus the tunneling current,  versus the
rotation angle  follows a $\sin^2$ behavior in contrast to the cosine
behavior of the  spin-polarized part of the tunneling current in the
SP-STM geometry when SOC is not included. 

\subsection{Crystal surfaces: $k_{||}$-Selection in STM \label{kparallel}}
The DOS in vacuum above the substrate can be computed with ab-initio 
method, but for bulk systems 
before getting the DOS in real space a summation over $k$--points 
in reciprocal space has to be performed within the Brillouin zone. It is 
instructive to realize that not all $k$-points contribute equally to the vacuum's DOS. Indeed 
on a surface probed by STM, because of symmetry reduction only the
parallel component, $\vec{k}_{||}$,   of the the three dimensional
Bloch vector  $\vec{k} = (k_{\perp},\vec{k_{||}})$ remains as a good
quantum number. Thus in vacuum the wave function,
$\psi_{\vec{k}_{||}\mu}$,  describing the surface  characterized on
the basis of Bloch theorem by a band index $\mu$ and  a wave-vector
$\vec{k}_{||}$  of the two-dimensional (2D) Brillouin zone is expanded
into basis functions:
\begin{equation}
\psi_{\vec{k}_{||}\mu}(\vec{r}_{||},z) =  \sum_{\alpha}
c_{\vec{k}_{||}\mu}^n d_{\vec{k}_{||}\mu}^n(z) \exp{[i(\vec{k}_{||}
  +\vec{G}_{||}^n)\vec{r}_{||}]}
\end{equation}
which are 2D plane waves parallel to the surface. $\vec{G}_{||}^n$
denotes the reciprocal  lattice vectors parallel to the surface while
$d^{\alpha}_{\vec{k}_{||}}$ are basis functions describing the decay of the
substrate's states into  vacuum and  can be obtained by solving the
one-dimensional Schr\"odinger equation in vacuum
 \begin{equation}
\left(-\frac{\hbar^2}{2m}\frac{d^2}{dz^2} + U(z) -\epsilon_{\mu} +
\frac{\hbar^2}{2m}(\vec{k}_{||}+\vec{G}_{||}^{\alpha})^2
\right)d^{\alpha}_{\vec{k}_{||}}(z) = 0,
\end{equation}
for a given vacuum potential $U(z)$ and reference energy
$\epsilon_{\mu}$.

In general, quantities possessing the crystal symmetry of the lattice
can be expanded into a set of symmetrized  functions. The local density
of states (LDOS)
\begin{eqnarray}
 n(\vec{r}_{||},z;\epsilon) &=&
 \sum_{\vec{k}_{||}\mu}{\delta(\epsilon-\epsilon_{\vec{k}_{||}\mu})
   |\psi_{\vec{k}_{||}\mu}(\vec{r}_{||},z)|^2}\\\nonumber
 &=&\sum_{\vec{k}_{||}\mu}\delta(\epsilon-\epsilon_{\vec{k}_{||}\mu})
 \sum_{\beta}n_{\vec{k}_{||}\mu}^{\beta}(z)\exp{(i\vec{G}_{||}^{\beta}\vec{r}_{||})}
\end{eqnarray}
where $\exp{i\vec{G}_{||}^{\beta}\vec{r}_{||}}$ are called star coefficients
and 
\begin{equation}
n_{\vec{k}_{||}\mu}^{\beta}(z) =
\sum_{\alpha\alpha'}c_{\vec{k}_{||}\mu}^{\alpha}c_{\vec{k}_{||}\mu}^{\alpha'*}
d_{\vec{k}_{||}\mu}^{\alpha}(z)d_{\vec{k}_{||}\mu}^{\alpha'*}(z)
\delta(\vec{G}^{\alpha}_{||}-\vec{G}^{\alpha'}_{||},\vec{G}^{\beta}_{||})
\end{equation}
$d$ could be  the $s$-orbital as proposed in TH model, or one can use
arbitrary orbitals according  to the derivative rule of Chen.

One could grasp the behavior of the LDOS in vacuum by considering for
simplicity that  $U(z) = 0$. Thus the exact solution of the one
dimensional Schr\"odinger equation in vacuum  gives $d_{\vec{k}_{||}}^{\alpha}
= \exp{(-\varkappa_{\vec{k}_{||}}^{\alpha} z)}$ where $z > 0$. The decay constant 
\begin{equation}
\varkappa_{\vec{k}_{||}}^{\alpha} =
\sqrt{2m|\epsilon_{\mu}|/\hbar^2+(\vec{k}_{||}+\vec{G}_{||}^{\alpha})^2}
\end{equation} 
and the related LDOS 
\begin{equation}
n_{\vec{k}_{||}\mu}^{\beta}(z) =
\sum_{\alpha\alpha'}c_{\vec{k}_{||}\mu}^{\alpha}c_{\vec{k}_{||}\mu}^{\alpha'*}
\exp{\left[-({\varkappa}_{\vec{k}_{||}\mu}^{\alpha}+{\varkappa}_{\vec{k}_{||}\mu}^{\alpha'})z\right]}
\delta(\vec{G}^{\alpha}_{||}-\vec{G}^{\alpha'}_{||},\vec{G}^{\beta}_{||})
\end{equation}
show obviously a strong dependence on $\vec{k}_{||}$ and on $\vec{G}_{||}^{\alpha}$.

The last equations demonstrate that the decay constant is the largest and
thereby the LDOS is the smallest  when  contributions of
$|\vec{k}_{||}+\vec{G}_{||}^{\alpha}|$   are significant. A decrease in $\varkappa$ of 0.1
\AA$^{-1}$  could lead to a reduction of the LDOS and of the tunneling
current of the order of 50\%.

\begin{figure}[ht!]
\begin{center}
\includegraphics*[width=1.\linewidth,angle=0,trim=0mm 120mm 0mm
  110mm]{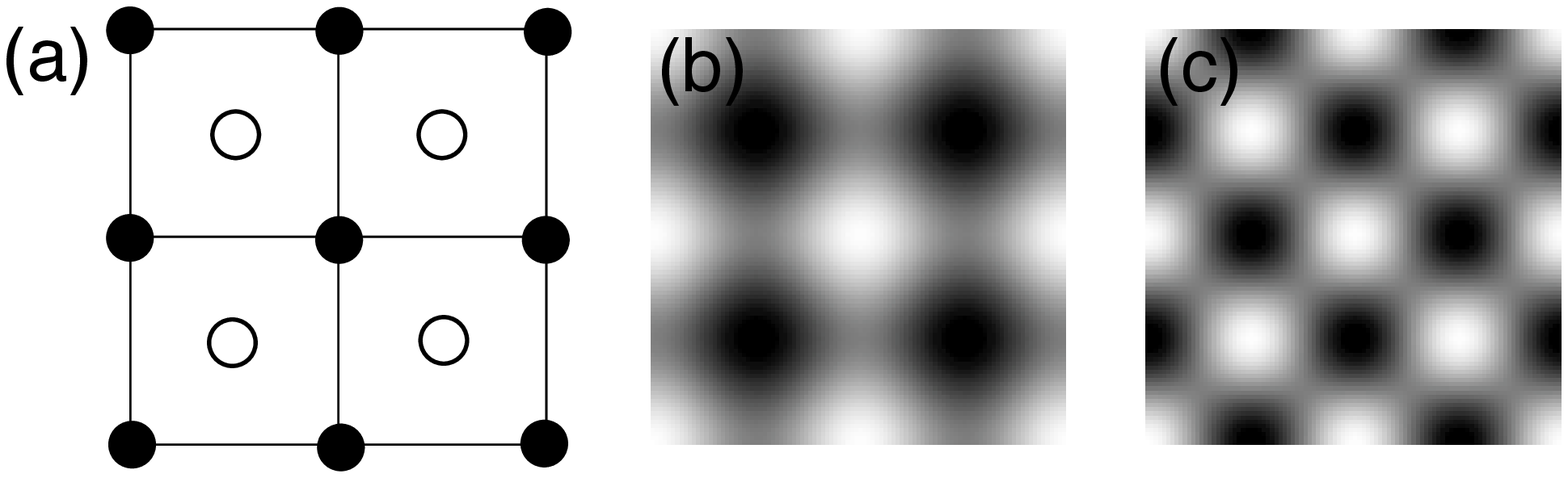} 
\caption{Star functions for a square lattice according to Ref.~\cite{Heinze_PhD}. 
(a) unit cell of a
  square lattice  with a two-atoms basis, i.e. representing a
  checkerboard structure. The first star function is a  constant
  contrary to the second one which leads to a chemical sensitivity on the
  two atoms (b). The  black atoms are displayed as protrusions and the
  white ones as depressions. Interestingly the third star function
   shows equally both atoms (c).  } 
\label{star-function}
\end{center}
\end{figure}
Also the nature of the lattice can affect tunneling, as demonstrated in 
the following example. We show the  first three vectors responding to the smallest
``stars'' (m=1, 2, 3) of reciprocal lattice vectors of  a square lattice
(see Fig.~\ref{star-function}), which would represent the bcc(001) or
fcc(001) surfaces.  The corresponding reciprocal lattice vectors are
$\vec{G}_{||}^{(1)} = (0,0)$,  $\vec{G}_{||}^{(2)} = (1,0)$, and
$\vec{G}_{||}^{(3)} = (1,1)$, expressed in units of $2\pi/a$, with
$a$ being the lattice constant and the star coefficients are:
\begin{eqnarray}
\exp{(i \vec{G}_{||}^{(1)} \vec{r}_{||} )} &=&
1\\ \exp{(i\vec{G}_{||}^{(2)}\vec{r}_{||})} &=&
\frac{1}{2}[\cos{(\vec{G}_{||,1}\vec{r}_{||})}  +
  \cos{(\vec{G}_{||,2}\vec{r}_{||})}
]\\ \exp{(i\vec{G}_{||}^{(3)}\vec{r}_{||})} &=&
\cos{[(\vec{G}_{||,1}+\vec{G}_{||,2})\vec{r}_{||}]}.
\end{eqnarray}
$\vec{G}_{||,1} = \frac{2\pi}{a}(1, 0)$ and $\vec{G}_{||,2} =
\frac{2\pi}{a}(0,1)$  are the two-dimensional reciprocal lattice
vectors.  In Fig.~\ref{star-function} the star functions are displayed
together with the 2D unit cell for a  checkerboard structure with two
different atom types. The first star function is a constant and
represents the lateral  constant part of the LDOS and thereby also of
the tunneling current in the TH-model.  Higher star coefficients can
contribute non-trivially to the final STM image.  The second
coefficient for example, allows to distinguish between the two kinds
of  atoms while the third coefficient does not,
i.e., chemical sensitivity  is probed only by the second star
coefficient. 

Any magnetic superstructure lowers the translational
symmetry. Therefore, smaller  reciprocal lattice vectors become
relevant for the spin-polarized part of the tunneling  current with
coefficients that are consequently exponentially larger than those of
the  unpolarized part.  


\section{Examples of simulations and experiments}

\subsection{Seeing the {Fermi surface} in real 
space via the induced charge oscillations}\iffindex{Fermi surface}
\iffindex{Friedel} {Friedel} oscillations define an important concept in quantum
mechanics. They are created after perturbing  an electron gas with an
impurity. Charge and magnetic oscillations  are then obtained in the
surrounding electron  gas. The shape and intensity  of these
oscillations contain important information on the impurity's
electronic structure and  on the band-structure of the host material
where the impurity is embedded. As mentioned earlier, STM allowed to
observe Friedel oscillations induced on surfaces  characterized by
two-dimensional electronic surface states. These surfaces, typically,
Cu(111), Ag(111) and Au(111) surfaces provided the right  playground
for experimental and theoretical investigations on charge variations induced
by impurities in a quasi-two-dimensional electron gas.   

\subsubsection*{Isotropic Friedel oscillations} 
 For Cu(111) the surface state shows a parabolic dispersion with a
 minimum at $\sim 0.5$ eV below the Fermi level. The corresponding band
 structure projected on the $\bar \Gamma - \bar M$ line of the
 2D-Brillouin zone is shown in Fig.~\ref{corrals}(a). The shaded
 regions indicate  the regions in $E-{\bf k}_\|$ space,  for which bulk
 eigenstates (Bloch waves) exist. Surface states can only exist in the
 white ''gap''-regions. Two such states are indicated.  Of special
 interest is the parabolic band with the minimum close to $E_F$, since
 this state is only partially occupied and gives rise to a
 two-dimensional metallic behavior, which is of great interest for
 the following.

\begin{figure}[ht!]
\begin{center}
\includegraphics*[width=.75\linewidth,angle=90,trim=50mm 0mm 0mm
  20mm]{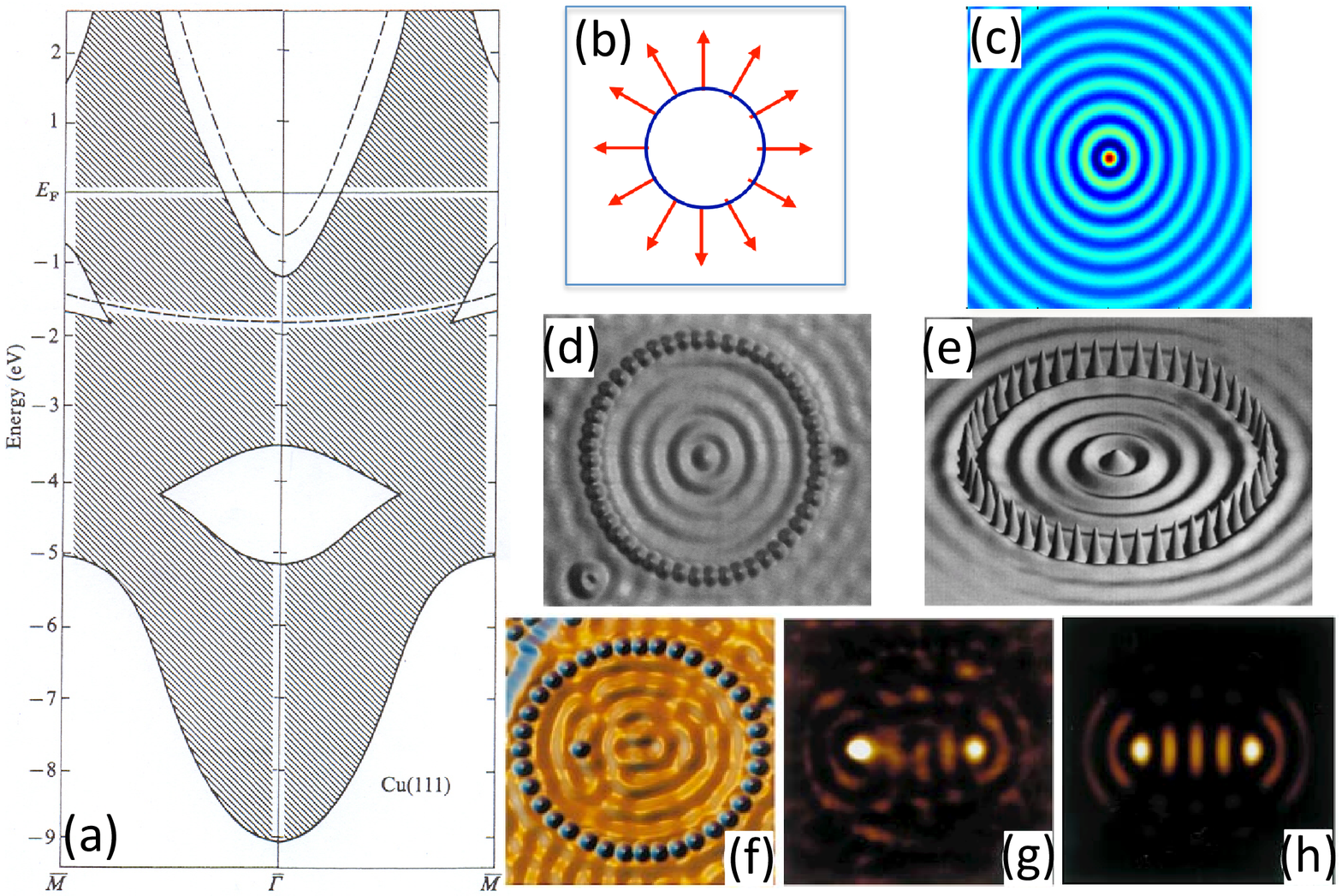} 
\caption{(a) Surface states (dashed curves) and bulk projected bands
  at a Cu(111) surface according to a six-layer surface band structure
  calculation~\cite{euceda}. In (b) and (c) illustration of the
  relation between the isotropic shape of the  Fermi surface (black
  contours), group velocities (red arrows) in (b) and the
  corresponding induced isotropic  charge oscillations in (c). In (d)
  and (e) is shown a comparison between the tunneling spectra obtained
  for a corral of Fe adatoms on Cu(111) surface as measured by Crommie
  {\it et al.}~\cite{crommie} (d) to the  calculations made with the KKR
  method~\cite{crampin}(e). Visualization of the quantum mirage with a
  mirage effect is shown in (f),  (g) and (h). {(f)} is a
  topography showing an ellipse with  $e = \frac{1}{2}$  and  a Co
  atom at the left  focus. In (g) the    associated $\frac{dI}{dV}$
  difference map shows the Kondo effect  projected to the empty right
  focus, resulting in a Co atom mirage. This experimental measurement
  compares well with the  calculated eigenmodes at $E_F$ (magnitude of
  the wave function is  plotted) as shown in (h).~\cite{manoharan} } 
\label{corrals}
\end{center}
\end{figure}

 For defects in the bulk, these Friedel oscillations of the charge
 perturbation vary for large distances $r$  as $1/r^3$ times an
 oscillatory function and are in the jellium model proportional to:

\begin{equation}
 \Delta n (r) \ \sim \ \frac{\cos (2k_F r+\delta)}{r^3}
\end{equation}

However in the case of adatoms on surfaces, the charge response decays
for long in-plane distances $r$ slower than in the bulk and is
determined by the surface states. In a free electron model, being well
suited for the above surface state for Cu(111), the charge density is
for large distances $r$ proportional to

\begin{equation}
 \Delta n (r) \ \sim \ \frac{\sin (2k_F r +\delta)}{r^{2}} 
\end{equation}

However, since also bulk states exist, which span most of the phase
space (see Fig.~\ref{corrals}(a)), the short range screening of the
defect is dominated by these states, while only the long ranged
behavior is determined by the surface state, which has a small wave
vector $k_F$ leading to long wave length oscillations.

In Fig.~\ref{corrals}(b) and (c), are shown the Fermi surface of a
two-dimensional electron gas with group velocities (vectors shown in red) 
and the  corresponding induced charge
around the impurity which is then isotropic and circular. The shape of
the induced  oscillations indicate that the related energy contour in
reciprocal space is circular.

Many authors have observed such long ranged oscillations around
adatoms, small clusters and steps on the Cu(111) surface in STM
experiments. Most prominent among these is the work of the team of 
Eigler {\it et
al.}~\cite{eigler,crommie,heller,manoharan}. 
By atomic manipulations they were able to construct
a corral \iffindex{Corral of adatoms} of Fe atoms on the (111) Cu surface, and have shown that the
surface states in the corral are more or less localized and form a
discrete spectrum of resonant states. As an illustration of these  we
show in Fig.~\ref{corrals}(d) and (e) a comparison of the experimental
measurements to the result of  calculations of Crampin {\it et
  al.}~\cite{crampin} obtained with the Korringa-Kohn-Rostoker Green
function (KKR) method   for a circular corral of 48 Fe atoms on the
Cu(111) surface. Shown are  the local density of states at the Fermi
energy at 5 \AA~ above the surface.  Within the corral one sees a
quantum well state with five maxima,  corresponding to a localized
state being more or less completely confined  to the corral. Outside
one sees oscillations arising from scattered  surface state electrons
at the corral, which decay with distance.

Let us shortly discuss the reason for the strong scattering of the
surface state electrons at the Fe atoms. Basically in the vacuum
region the  full potential of Fe acts as a scattering center for the
surface wave,  being much stronger than the scattering at an Fe
impurity in the bulk,  where only the change of the Fe potential with
respect to the host potential  is effective. Moreover the wave vector
$k_F$ is relatively small, such that  the wave length is considerably
larger than the spacing between the Fe atoms.  Therefore the surface
wave does not ``see'' the corrugation of the Fe ring  and is strongly
reflected as in a cylindrical well. In fact the sequence of  resonances
can be well described by such a quantum well model, as has been shown
recently~\cite{szunyogh}. The most fascinating corral experiments are
the observation of atomic mirages in an elliptical quantum
well~\cite{manoharan}. An  ellipse has the well known property that all
classical waves emanating from one  of the two focus points in every
direction are reflected from the ellipse wall  and focused in the
second point, where these waves add up coherently  since each such
partial wave has the same path length and therefore the same  phase
shift. This is illustrated in Figs.~\ref{corrals}(f--h) taken from 
Ref.~\cite{manoharan}. Fig.~\ref{corrals}(f)
shows  the STM topography for an ellipse   with a given  eccentricity,
including one Co atom at the left focus point.  Fig.~\ref{corrals}(g)
shows the $\frac{dI}{dV}$  difference maps,  i.e. the change  of
the STM intensity map with respect to a small bias voltage V, which
corresponds  in the calculations to the local density of states in the
vacuum region  at the height of the STM tip. We see clearly two
intensity spots, the real  Co atom at the left focus and its image at
the right focus. Thus in the  empty focus we see the same accumulation
of charge in the surface state as around  the Co atom; therefore the
image is called a quantum mirage. In fact the  Co atom is a Kondo
impurity and a strong and sharp Kondo peak appears  only in a very
small energy region of about 10 meV around the Fermi level. Moreover
the large mirage only appears, if one of the quantum well states falls
into this  energy region. Fig.~\ref{corrals}(h) shows the  calculated
localized eigenstate observed in the  experiment. The calculated local
density of states compares very well with the  $\frac{dI}{dV}$ curve
shown in Fig.~\ref{corrals}(h). Thus  several conditions have to be
satisfied for the Co mirage to appear:(i) the Co-atom has to sit in  a
focus point; if it sits at another position away from the focus point,
no  image appears, (ii) the bias voltage has to be such, that it
coincides with an  eigenstate of the ellipsoidal corral having maxima
at the focus points, (iii)  finally the image is particularly intense,
if the eigenvalue coincides  with the Kondo resonance. This concept
has been recently extended theoretically by Stepanyuk and co-workers
for  the induced magnetization confined in magnetic corrals~\cite{Stepanyuk}.

\subsubsection*{Focusing effect}\iffindex{Focusing effect}
If one manages to embed a Co-impurity few layers underneath the
surface and try to visualize the induced Friedel oscillations on the
surface, strange patterns  are observed.  Recently, we have shown by
ab-initio calculations combined with STM observations that anisotropic
localized oscillations can be observed on top of Cu(111) and  Cu(001)
surfaces due to the presence of buried Co 
impurities~\cite{Lounis_focusing,Lounis_focusing2}. These
anisotropic ripples show that the usual isotropic free-electron model
is not valid in such  real situations. We demonstrated that these
intriguing features are nothing else than a visualization in real
space of parts of the bulk copper Fermi surface that are relatively
flat.  For the comparison between theory and STM, use is made of the
Tersoff-Hamann model stating, as mentioned earlier, that scanning
tunneling spectra can be related to the DOS in a certain energy
interval in the vacuum.
\begin{figure}[ht!]
\begin{center}
\includegraphics*[width=1.\linewidth,angle=0,trim=10mm 230mm 20mm
  0mm]{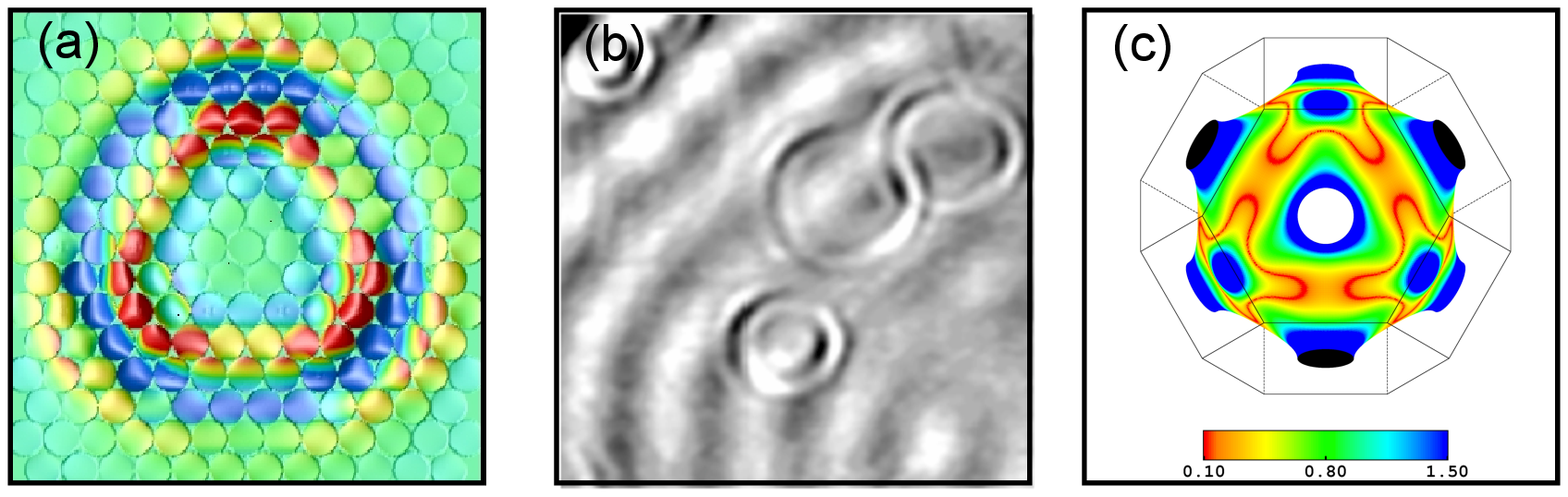} 
\caption{(a) Impurity induced charge density around $E_F$ for an area
  of $\approx 30 \times 30$ \AA$^2$~ calculated  at a height $\approx 6.1$ 
  \AA~ above the Cu(111) surface with a Co impurity sitting in the
  6$^{th}$ layer below the surface. (b) Experimental  STM topographies
  for an area of $90 \times 90$ $\AA^2$~ ($-80 mV, 1 nA$) of four Co atoms below
  the Cu(111) surface. (c) Fermi surface of copper represented along
  the (111) direction. The inverse mass tensor corresponding to the
  denominator of Eq.~\ref{eq.focusing} is represented by the color in units of the
  inverse electron mass. Small values represented in red lead to high
  intensities of the charge variation.} 
\label{compilation_focusing}
\end{center}
\end{figure}

Fig.~\ref{compilation_focusing}(a) shows an example of the results of our
simulations: the case of a Co impurity sitting at the $6^{th}$ layer
below the Cu(111) surface (-12.5\AA~below the surface). The charge
induced in the vacuum has been computed up to an area of $30 \times 30$
\AA$^2$  above the impurity. One notices the triangular shape of the
induced charge with high values at the  corners of the triangle. In
addition, a one and a half period ripple can be observed to oscillate
from the red positive value to the  blue negative values and finally
to the almost zero green values. The same period was also noticed in
the STM experiment (Fig.~\ref{compilation_focusing}(b)). 
To understand such a phenomenon, we start by
giving the  form of the induced variation of the DOS in the vacuum at
position 0 by some buried Co impurity sitting at a position defined by
$\vec{R}$ and inducing a change in the potential $\Delta V$:
\begin{equation}
\Delta {n}_{\mathrm{vacuum}} (\epsilon)\sim -\frac{1}{\pi}\Im\left[\int\int
d^3rd^3r'
G_0(0,\vec{r};\epsilon) \Delta V(\vec{r}) G_0(\vec{r},0;\epsilon)\right]  
\end{equation}
where the Green function, $G_0$, obtained from the KKR method, describes the pure substrate.  At
very large distances $R$ between the impurity and  the vacuum site one
can apply the stationary phase approximation and end up with a result
similar to that of the well-known theory of interlayer exchange
coupling:
\begin{equation}
R^2 \Delta \mathrm{n}_{\mathrm{vacuum}} \propto \frac{1}{\left|
  \frac{d^2E}{dk_x^2} \cdot \frac{d^2E}{dk_y^2}\right|}.
\label{eq.focusing}
\end{equation}
The denominator of this equation is a measure of the curvature of the
constant energy surface, i.e., the shape of the constant energy
surface affects the propagation of the electrons.  Additionally, one
can show that the electronic waves are directed by the
group velocities. Since, states at the Fermi energy ($E_F$) are probed
experimentally, the constant  energy surface corresponds to the Fermi
surface: a small value of the curvature means that the Fermi surface
has a flat region leading to large values of the DOS and to strong
focusing of intensity in this space region determined by the group
velocity. In Fig.~\ref{compilation_focusing}(c), we show  the Fermi surface, computed with
ab-initio, of Cu oriented with the (111)-neck direction normal to the
drawing plane.  The Fermi surface is colored following the strength of
the denominator of the right hand side of Eq.~\ref{eq.focusing}.
  The shape of the low values of this denominator,
corresponding to the flat regions of the Fermi surface is  found to be
rather triangular along the (111) direction in accordance with our
simulations  of the induced Friedel oscillations. One can understand
that the flat region seen within the triangle in
Figs.~\ref{compilation_focusing}(a) and (b) is  induced by the neck of the Cu Fermi surface
along the (111) direction that does not allow electrons to propagate.

 Once the shape of
the propagator $G$ is known, either by STM-investigation  or
calculations based on density functional theory many additional effects can be
predicted. The strong directionality of electron propagation even for a simple
metal such as Cu, has consequences in many fields. For example, the spatially anisotropic characteristics
should also be equally present  in the RKKY interactions between
magnetic impurities. Indeed we have shown that the interaction between
Co adatoms on Pt(111) surface or  Fe adatoms on Cu(111) surface is
very anisotropic~\cite{Wiebe1,Wiebe2},  which is obviously induced by the same physics
discussed in this subsection, i.e. the anisotropic shape of the Fermi
surface. 

\subsection{Magnetism on surfaces with SP-STM}
An example calculated by Heinze and collaborators~\cite{heinze2002} is shown in
Fig.~\ref{sp-stm-results}(a) and (b).  A Cr overlayer on Ag(111) shows
a row-wise antiferromagnetic structure (actually the ground state
calculated by the authors is non-collinear, but for now we focus on
the  antiferromagnetic state). The row-wise antiferromagnetism is
indicated by the arrows drawn  at the atomic positions, showing the
magnetization direction of each atom. A scan of the surface  with a
non-magnetic STM tip will show the chemical unit cell
(Fig.~\ref{sp-stm-results}(a)):  bright spots correspond to the Cr
atoms, from which a tunneling current flows to the tip when the apex
atom is above  them. But a scan with a spin polarized tip reveals the
magnetic structure (Fig.~\ref{sp-stm-results}(b)).  Instead of bright
spots appearing around each atom, now bright stripes emerge at the
rows with a magnetization direction parallel to the one of the tip,
while  dark stripes appear at the rows with opposite magnetization.

In these calculations, the assumption of a fully spin-polarized tip
was made: i.e. $n_{T\downarrow}(E_F) = 0$.  In Figs.~\ref{sp-stm-results}(a) and 
(b), we see
that the bright stripes (high current) appear when the substrate
magnetization  is parallel to the one of the tip. It should be noted
that this is by no means guaranteed for all cases; it can  well be
that the antiparallel orientation favors  the tunneling
current. Eventually a case might be encountered where the parallel and
antiparallel  configurations give an almost equal  signal at $E_F$. In
this case spectroscopy is a very valuable tool, since one can detect
the  signal at other energies, choosing a voltage for which the two  spin
directions give considerably different results.

\begin{figure}[ht!]
\begin{center}
\includegraphics*[width=.6\linewidth,angle=90,trim=40mm 0mm 0mm
  30mm]{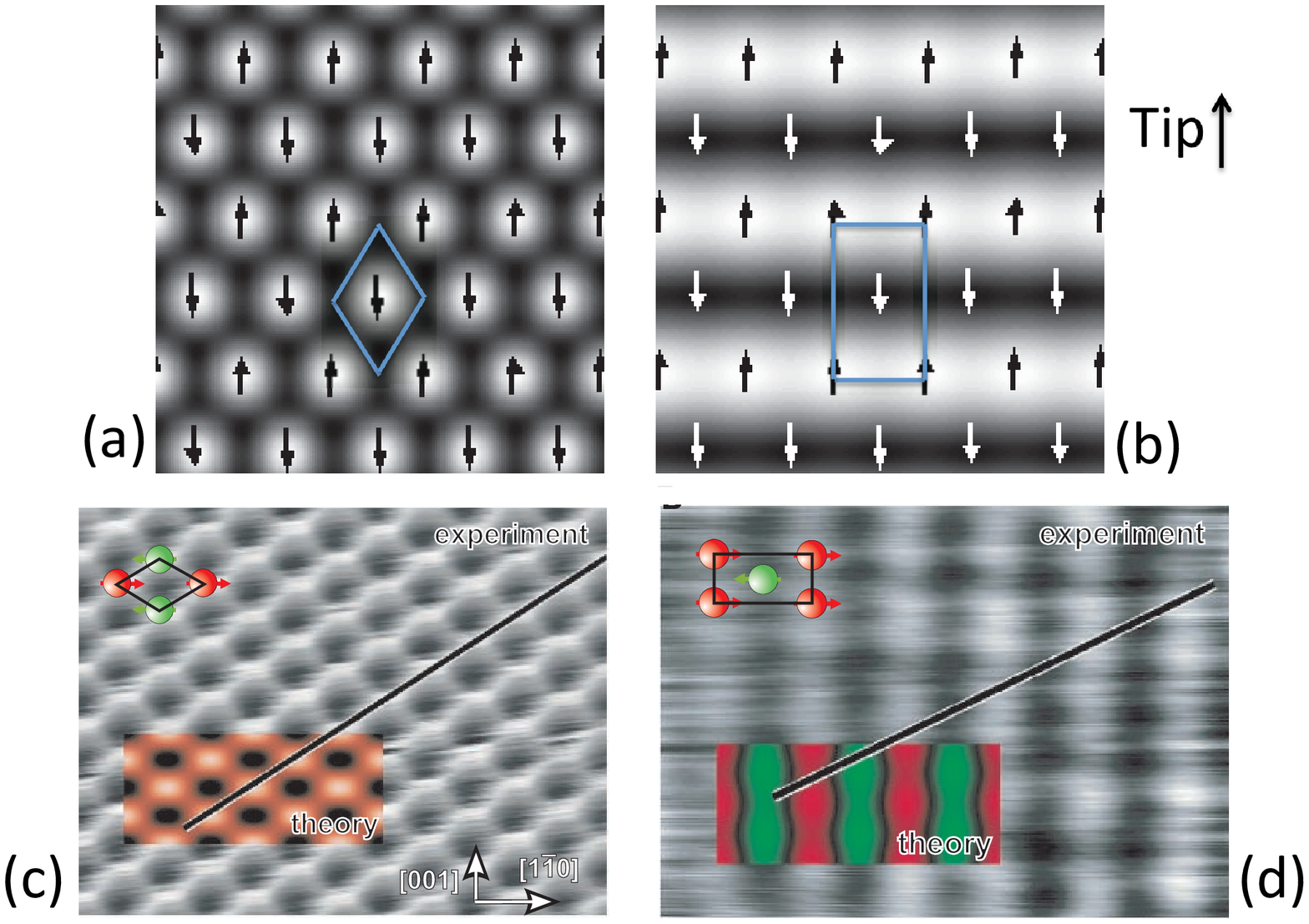} 
\caption{Calculated STM picture of an antiferromagnetic Cr monolayer
  on Ag(111) fcc surface after Ref.~\cite{heinze2002} and Mn  monolayer on W(110)
  surface after Ref.~\cite{Heinze2000}. In  (a) shows the result obtained
  assuming a non spin-polarized tip, revealing the chemical unit-cell
  (drawn parallelogram); Cr atoms appear as white filled circles. In
  (b) a spin-polarized tip with magnetization direction  in-plane as
  indicated. In this case the magnetic unit cell emerges (drawn as a
  rectangle), giving alternating  black and white stripes. The drawn
  arrows indicate the magnetization direction of each Cr atom.  A
  comparison of SP-STM measurements and first-principles calculations
  for the case of Mn layer on W(110) is shown in (c)  considering a
  non-magnetic tip and in (d) when the tip is magnetic. Once more the
  unit-cell of the calculated magnetic  ground state is shown in the
  insets of (c) and (d).} 
\label{sp-stm-results}
\end{center}
\end{figure}

Another example of comparison between experiment and theory is shown
in  Figs.~\ref{sp-stm-results}(c) and (d) for the case of Mn overlayer
on W(110) surface.  This work, realized by Heinze {\it et al.}~\cite{Heinze2000}, 
was the first observation of antiferromagnetism in a single magnetic layer. 
  They found, as in the precedent example,
that non–spin-polarized tunneling electrons image the chemical
surface unit  cell without any magnetic contribution, whereas
spin-polarized electrons probe the change in translational symmetry
due to the magnetic superstructure, which gives rise to a different
image corresponding  to the respective magnetic structure. The magnetic ground
state that is antiferromagnetic in a  checkerboard arrangement of Mn
atoms with magnetic moments of opposite direction  and an easy axis of
the magnetization oriented in  the film plane leads in the theoretical
STM image to a stripe pattern  similar to the one obtained
experimentally. However, when the STM  tip is considered non-magnetic,
all Mn atoms become equivalent in the chemical unit cell and the
STM-pattern  becomes diamond-like both experimentally and
theoretically. A detailed discussion on magnetism and in particular on magnetism at 
surfaces is given in Chapter {\bf C 4} by S.~Bl\"ugel.

\subsection{Magnetic domain walls with TAMR-STM}

The first experimental verification of the dependence of the DOS on 
the magnetization orientation was 
provided experimentally by Bode  {\it et al.}~\cite{Bode2002} using a non spin-polarized 
STM on Fe double layers deposited on W(110) substrate. This effect discussed in 
Section~\ref{TAMR}, 
was already predicted theoretically~\cite{Lessard}. The substrate chosen by Bode  {\it et al.} is 
well known for having a nanometer-scale domain structure. In other words, the magnetization of the sample rotates 
at the nanometer-scale as shown in Fig.~\ref{compilation_TAMR}(a) and the idea is to probe the rotation of the magnetization  
at different positions using the spectroscopic mode, i.e. to measure the $dI/dV$ spectra. 
\begin{figure}[ht!]
\begin{center}
\includegraphics*[width=.8\linewidth,angle=90,trim=0mm 0mm 0mm
  20mm]{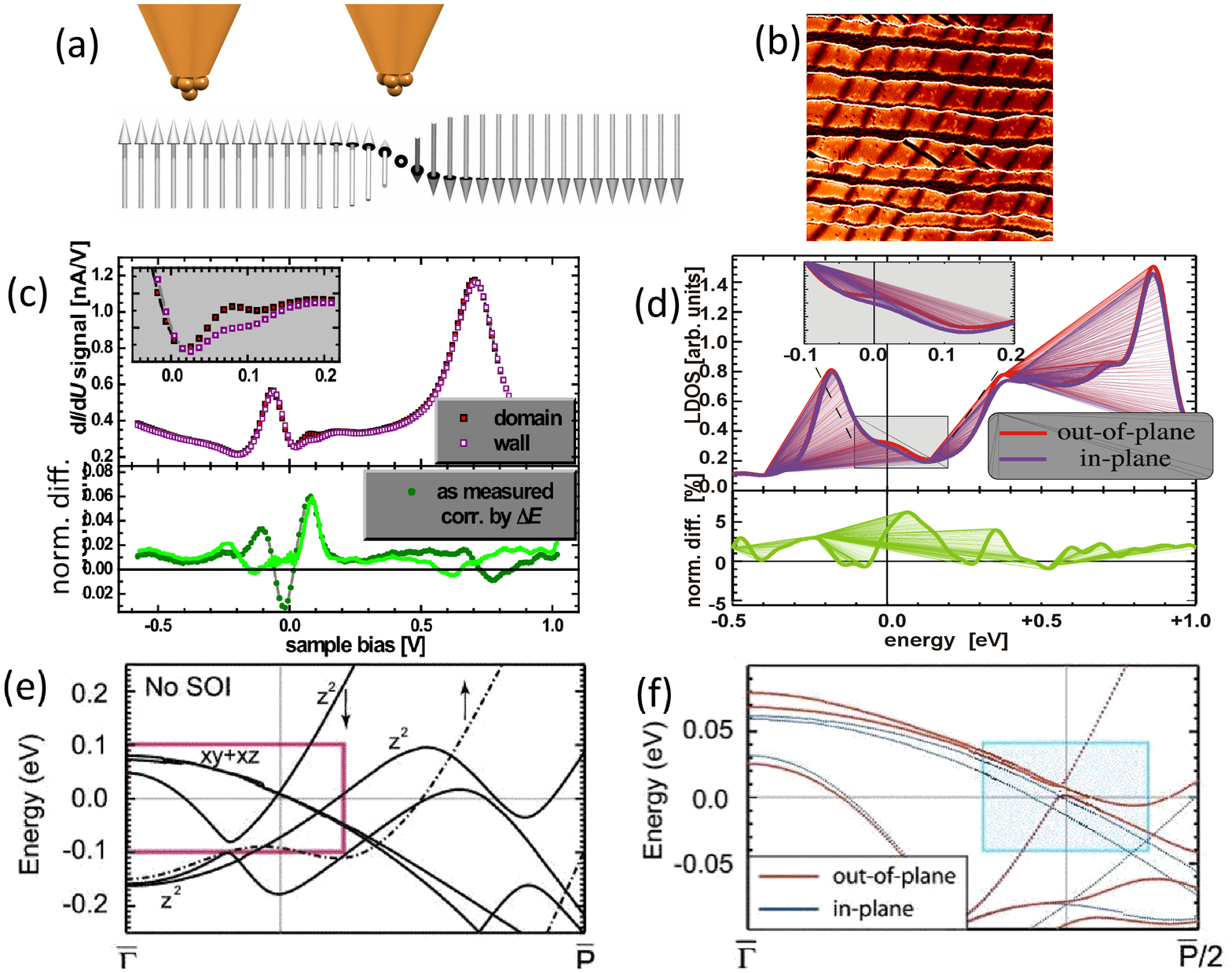} 
\caption{(a) Measurement of a domain wall with an STM-probe at different location: within the domain and at the domain-wall. 
(b) shows an image of the $dI/dV$ signal measured on two-monolayers Fe deposited on W(110) substrate. Stripes of $\sim$ 50 nm widths can be 
observed with bright (within the domain $\sim$ in-plane orientation of the magnetization) and dark (at the domain-wall $\sim$ out-of-plane orientation of 
the magnetization) regions. In (c) and (d) are shown the bias-dependence of the $dI/dV$ signal and the theoretical LDOS at different locations: 
domain (in-plane magnetization) and domain-wall (out-of-plane) regions. A TAMR ratio is computed and shown in lower parts of (c) and (d). The band-structure 
calculated without (e) and with SOC (f) demonstrates the impact of SOC and of the orientation of the magnetization on the 
electronic structure of the sample. Figures taken and adapted from Ref.\cite{Bode2002}} 
\label{compilation_TAMR}
\end{center}
\end{figure}

Interestingly, the domain walls are visible with a non-magnetic W tip (see Fig.~\ref{compilation_TAMR}(b)) along the different 
stripes propagating lateraly: 
At the position of the domain wall the differential conductivity $dI/dV$ is reduced with respect to the domain. 
As revealed by the local tunneling spectra (Fig.~\ref{compilation_TAMR}(c)), 
this contrast is caused by a tiny difference which is energetically located just above 
$E_F$ (see inset):
 while the $dI/dV$ spectrum measured with the tip positioned above the domain exhibits a weak peak at a bias of 0.07 eV, 
this peak is almost absent in the domain wall spectra. This is further illustrated in the lower part of Fig.~\ref{compilation_TAMR}(c) 
by the plot of the TAMR ratio calculated as 
$[(dI/dV)_{\mathrm{D}} - (dI/dV)_{\mathrm{W}}] /  [(dI/dV)_{\mathrm{D}} + (dI/dV)_{\mathrm{W}}]$, where 
the $dI/dV$ are taken within the domain (D) or at the domaine wall (W).\footnote{If the ratio 
is calculated on the original data (as measured), a pronounced oscillation can be found just below $E_F$. This oscillation is not caused by any additional or missing spectroscopic 
features in the domain wall of the $dI/dV$ spectrum with respect to the spectrum measured at domains but by an overall 
energetic shift $\sim 11$ meV. The physical origin is different work functions in domains and domain walls. After correction, the oscillation 
below $E_F$ has almost perfectly disappeared.} For comparison,  
first-principles calculations of the LDOS and TAMR ratio on the same system including SOC 
were performed considering different rotations (out-of-plane and in-plane) 
of the magnetic moments are shown in Fig.~\ref{compilation_TAMR}(d). The agreement is good and it is found theoretically that 
the two minority-spin $d_{z^2}$ states at the Fe surface lead to 
pronounced peaks at -0.18 and +0.85 eV and can be identified with the experimental peaks at -0.08 eV and +0.7 eV.  
A closer look (cf. inset in Fig.~\ref{compilation_TAMR}(d)) reveals a significant enhancement 
of the LDOS for the out-of-plane magnetized film within an energy interval of about 100 meV above $E_F$.
The theoretical TAMR signal is qualitatively in line with the experimental signal: The peak at -0.3 eV could be identified with the 
experimental peak at -0.24 eV. The origin of the dependence of the LDOS on SOC can be traced back to the change of 
the band-structure once the magnetization is rotated as discussed in Section~\ref{TAMR}. 
 The band-structure without  and with 
SOC are plotted in Figs.~\ref{compilation_TAMR}(e) and (f). States at the $\overline{\Gamma}$ point decay the lowest into vacuum (as discussed in 
Section~\ref{kparallel}). Around -0.18 eV and +0.05 eV, the $d_{z^2}$-band crosses the $\overline{\Gamma}$-point.  
Because of their appropriate extension in the $z$-direction, the $d_{z^2}$ 
orbitals decay slower 
into the vacuum compared to the $d_{xz}$ and $d_{yz}$ orbitals. Interestingly, the crossing of the minority $d$ bands 
that occured without SOC is avoided once SOC is included and if the magnetization is rotated out-of-plane (red line in  Fig.~\ref{compilation_TAMR}(f)).  
This has a strong impact by shifting, for example, the crossing of the bands at the $\overline{\Gamma}$-point. Overall such a small effect is enough 
to induce a change in the LDOS that is observable with STM. We note that 
the use of TAMR within STM was recently extended to probe adatoms deposited on surfaces with magnetic domain walls~\cite{Neel,Serrate}.

\section*{Acknowledgment}
I would like to acknowledge the HGF-YIG Programme VH-NG-717 
(Functional Nanoscale Structure and Probe Simulation Laboratory--Funsilab) for financial support.


\end{document}